\documentclass[
aip,
 amsmath,amssymb,
 reprint,%
]{revtex4-1}

\usepackage{graphicx}% Include figure files
\usepackage{dcolumn}% Align table columns on decimal point
\usepackage{bm}% bold math
\usepackage[utf8]{inputenc}
\usepackage[T1]{fontenc}
\usepackage{mathptmx}
\usepackage{etoolbox}
\usepackage{xcolor}
\usepackage[normalem]{ulem}

\makeatletter
\def\@email#1#2{%
 \endgroup
 \patchcmd{\titleblock@produce}
  {\frontmatter@RRAPformat}
  {\frontmatter@RRAPformat{\produce@RRAP{*#1\href{mailto:#2}{#2}}}\frontmatter@RRAPformat}
  {}{}
}%
\makeatother
\begin{document}

\preprint{AIP/123-QED}

\title{Selecting Relevant Structural Features for Glassy Dynamics by Information Imbalance}

\author{Anand Sharma}
\affiliation{Indian Institute of Science Education and Research,
Dr. Homi Bhabha Road, Pashan, Pune 411008, India}

\affiliation{Univ. Grenoble Alpes, CNRS, LIPhy, 38000 Grenoble, France}

\author{Chen Liu}
\affiliation{Innovation and Research Division, Ge-Room Inc., 93160 Noisy le Grand, France}

\author{Misaki Ozawa*}
\affiliation{Univ. Grenoble Alpes, CNRS, LIPhy, 38000 Grenoble, France}

\email{* misaki.ozawa@univ-grenoble-alpes.fr}

\date{\today}% It is always \today, today,

\begin{abstract}

We investigate numerically the identification of relevant structural features that contribute to the dynamical heterogeneity in a model glass-forming liquid. By employing the recently proposed information imbalance technique, we select these features from a range of physically motivated descriptors. This selection process is performed in a supervised manner (using both dynamical and structural data) and an unsupervised manner (using only structural data). We then apply the selected features to predict future dynamics using a machine learning technique. 
One of the advantages of the information imbalance technique is that it does not assume any model {\it a priori}, i.e., it is a non-parametric method.
Finally, we discuss the potential applications of this approach in identifying the dominant mechanisms governing the glassy slow dynamics.

\end{abstract}

\maketitle

\section{Introduction}

Glass-forming liquids exhibit a significant slowing down in dynamics as the temperature decreases. Relaxation times and transport coefficients, like viscosity, increase by more than ten orders of magnitude within a small temperature range~\cite{berthier2011theoretical,binder2011glassy,ediger1996supercooled}.
In addition to growing timescales, glass-forming liquids demonstrate heterogeneous spatial patterns of dynamics composed of mobile and immobile domains, called dynamical heterogeneity, which is another hallmark of glassy dynamics~\cite{berthier2011dynamical,karmakar2014growing}. The magnitude of dynamical heterogeneity increases with decreasing temperature, implying the emergence of highly non-trivial spacial correlations at lower temperatures. 

It has been reported that such heterogeneous dynamical patterns are statistically reproducible if one looks at the mobility or dynamic propensity of each particle  obtained from an ensemble of many independent trajectories starting from a given static configuration, called the iso-configurational ensemble~\cite{widmer2006predicting,berthier2007structure,widmer2008irreversible}. Such numerical observations suggest the existence of a relationship between the initial structure and future dynamics. Frustration-based theories attribute glassy dynamical behaviors to the existence of domains with locally preferred structures~\cite{tarjus2005frustration}. Indeed, extensive numerical and experimental works revealed that locally preferred structures are highly correlated with slow dynamics and dynamical heterogeneities~\cite{coslovich2007understanding,patrick2008direct,hocky2014correlation,jack2014information,turci2017nonequilibrium} (see e.g., Ref.~\onlinecite{royall2015role,tanaka2019revealing} for review).

Besides locally preferred structures, various physically-motivated descriptors characterizing local structural environments (e.g., local energy and density~\cite{widmer2005relationship,jack2014information}, geometrical order parameters~\cite{tanaka2010critical,tong2018revealing}, mesoscopic elastic stiffness~\cite{kapteijns2021does}, local entropy~\cite{berthier2021self}, etc.) have been proposed and tested to find a relation between structure and dynamics. 
Some of the proposed descriptors correlate well with future dynamics to some extent. While some descriptors have apparently different physical origins, some others share essentially the same information. Besides, recent studies have highlighted the role of coarse-graining on descriptors~\cite{tong2018revealing,boattini2021averaging}. Coarse-graining with an optimal length scale enhances the correlation between dynamics and averaged descriptors.
These extensive investigations lead to a question: What are the relevant (or main) physical descriptors responsible for glassy dynamics among a pool of many proposed descriptors with various coarse-graining lengths?
This question is also related to identifying the correct theory of glass transition among numerous proposals based on distinct physical mechanisms~\cite{tarjus2011overview}. Typically, each theoretical scenario attributes a specific quantity to glassy dynamics, such as entropy~\cite{berthier2019configurational,sengupta2012adam} in the random first-order transition theory~\cite{lubchenko2007theory,bouchaud2004adam}, elasticity~\cite{kapteijns2021does} in the solid-that-flow picture~\cite{dyre2024solid}, and so on. Thus, selecting the main physical descriptor(s) relevant to glassy dynamics would shed light on this fundamental theoretical issue.

Recently, machine learning techniques have been applied to reveal the structure-dynamics relationship from another angle.
Machine learning techniques specific to this problem consist of two approaches, supervised and unsupervised learning. Supervised learning in this context employs both structure (configurations, structural descriptors) and dynamics (propensity, mobility) data in training and forecasts the future dynamics of each particle of an unseen configuration~\cite{cubuk2015identifying,schoenholz2016structural,bapst2020unveiling,jung2023predicting,shiba2023botan,pezzicoli2022se}. 
Moreover, very recent studies have identified important descriptors for aged~\cite{ciarella2023finding,janzen2024classifying} and active~\cite{janzen2023dead} glasses using post-hoc explainable machine learning methods.
Unsupervised learning, instead, does not require dynamics data in training~\cite{boattini2020autonomously,oyama2023deep,coslovich2022dimensionality}. It consists of dimensional reduction of structural data and their clustering. A clustered map based solely on structural information is compared with a dynamical propensity map, and their correlations are evaluated~\cite{coslovich2022dimensionality}. 

In the language of machine learning and data science, identifying relevant features (descriptors in our case) is referred to as {\it feature selection}. 
%Several feature selection methods have been proposed.
One can easily perceive that the feature selection process is reduced to comparing two features, say $X^{(1)}$ and $X^{(2)}$, and asking whether one is more informative than the other.  By using the standard statistical measures such as the Pearson coefficient $\rho$, one can perform this comparison by computing $\rho$ between $X^{(1)}$ (or $X^{(2)}$) and $Y$, denoted as, $\rho_{X^{(1)}, \ Y}$ (or $\rho_{X^{(2)}, \ Y}$), where $Y$ is the desired output data (dynamic propensity in our case). 
Based on the values of $\rho_{X^{(1)}, \ Y}$ and $\rho_{X^{(2)}, \ Y}$, one can conclude which variable, $X^{(1)}$ or $X^{(2)}$, is more relevant to $Y$.
However, when the output data $Y$ is not available (the case of unsupervised learning), it is hard to assess whether one is more informative than the other by using solely $X^{(1)}$ and $X^{(2)}$, because $\rho_{X^{(1)}, \ X^{(2)}}$ is {\it symmetric} with respect to exchanging $X^{(1)}$ and $X^{(2)}$ by construction, i.e., $\rho_{X^{(1)}, \ X^{(2)}}=\rho_{X^{(2)}, \ X^{(1)}}$.
This is also the case for other statistical measures such as the mutual information $I$, i.e., $I(X^{(1)}, \ X^{(2)})=I(X^{(2)}, \ X^{(1)})$.
%Thus, one requires a sort of information measure between $X^{(1)}$ and $X^{(2)}$ constructed in an asymmetric way.

Recently, Ref.~\onlinecite{glielmo2022ranking} introduced a new information measure, called information imbalance, $\Delta(A \to B)$. It is defined in an {\it asymmetric} way to assess how much $A$ predicts (or determines) $B$.
By comparing $\Delta(A \to B)$ and $\Delta(B \to A)$, one can ask if $A$ and $B$ are equivalent, independent, or if
one is more informative than the other.
In particular, this new method was applied to selection of relevant policies on the Covid-19 epidemic state relevant for the number of deaths in the future~\cite{glielmo2022ranking}. Besides, it was applied to dimensional reduction of structural descriptors for atomic systems~\cite{glielmo2022ranking}.
This pioneering work is followed by
Ref.~\onlinecite{del2024robust}, where they used the information balance techniques to detect causality in simulated dynamical systems and real-world data taken from the electrophysiological signals of human brains.
In Refs.~\onlinecite{donkor2023machine,donkor2024beyond}, the information imbalance analysis is used to assess the predictivity and interpretability of various descriptors for systems of water molecules.
In addition to scientific applications, Ref.~\onlinecite{wild2024maximally} conducted feature selection from clinical databases to obtain a maximally informative set of features relevant to the severity of the disease.

In this paper, we employ information imbalance techniques to select structural features relevant to glassy slow dynamics. 
In particular, we select several relevant features from a pool of physically motivated descriptors composed of local energy, local number density, local volume fraction, coordination number~\cite{tong2023emerging}, bond-orientational order~\cite{kawasaki2007correlation}, and steric bond order~\cite{tong2018revealing}, with various coarse-grained lengthscales. We perform feature selection in both supervised (using dynamic propensity) and unsupervised (without dynamical data) settings.

First, we analyze the correlation between dynamic propensity and various physical descriptors varying the coarse-grained length scale using the conventional Pearson coefficient. We then confirm that the information imbalance reproduces the general trend obtained by the Pearson coefficient. Next we select relevant features in a supervised way by minimizing the information imbalance between descriptors and dynamic propensity. We also perform feature selection in an unsupervised way using only structural data. Finally, we use the selected features (from supervised and unsupervised manners) as input for the prediction of dynamics by machine learning and compare performance among different feature selection methods.

This paper is organized as follows.
Section~\ref{sec:Methods} describes our simulation model, physically motivated descriptors, and a concise introduction to the information imbalance techniques. 
We then present in Sec.~\ref{sec:Results} the results for the comparison between the Pearson coefficient and the information imbalance, feature selections by both supervised and unsupervised learning, and their application to machine learning dynamic propensity.
We finally discuss future perspectives in Sec.~\ref{sec:discusssion_conclusion}.

\section{Methods}
\label{sec:Methods}

\subsection{Simulation model}

We study a three-component glass-forming liquid model composed of small (S), medium (M), and large (L) particles in two-dimensions. We add an additional species (M) to a binary model (composed of S and L species) studied widely~\cite{falk1998dynamics,barbot2018local,barbot2018local,lerbinger2022relevance}.

Interaction between two particles is given by the Lennard-Jones potential, $v_{\alpha \beta}(r)=4 \epsilon_{\alpha \beta} \left[(\sigma_{\alpha \beta}/r)^{12} - (\sigma_{\alpha \beta}/r)^{6}\right]$, where $\alpha, \beta={\rm S}, {\rm M}, {\rm L}$. The potentials are modified to be twice
continuously differentiable functions at a cutoff, following Ref.~\onlinecite{barbot2018local}.
The parameters, $\sigma_{\alpha \beta}$ and $\epsilon_{\alpha \beta}$, for the potentials are given as follows:
$\sigma_{\rm LL}=2 \sin(\pi/5)\simeq 1.18$, $\sigma_{\rm SS}=2 \sin(\pi/10) \simeq 0.62$, $\sigma_{\rm LS}=1$,
$\sigma_{\rm LM}=(\sigma_{\rm LL}+\sigma_{\rm LS})/2$,
$\sigma_{\rm MS}=(\sigma_{\rm LS}+\sigma_{\rm SS})/2$, and
$\sigma_{\rm MM}=(\sigma_{\rm LL}+\sigma_{\rm SS})/2$.
$\epsilon_{\rm LL}=1/2$, 
$\epsilon_{\rm SS}=1/2$, 
$\epsilon_{\rm LS}=1$,
$\epsilon_{\rm LM}=(\epsilon_{\rm LL}+\epsilon_{\rm LS})/2$,
$\epsilon_{\rm MS}=(\epsilon_{\rm LS}+\epsilon_{\rm SS})/2$,
and
$\epsilon_{\rm MM}=(\epsilon_{\rm LL}+\epsilon_{\rm SS})/2$.
All particles have the same mass, which is set to $1$.
The number of particles is $N=N_{\rm S}+N_{\rm M}+N_{\rm L}=4000$, where $N_{\rm S}=1760$, $N_{\rm M}=800$, and $N_{\rm L}=1440$ are the number of particles for small (S), medium (M), and large (L) particles, respectively.
We consider $NVT$ canonical ensemble. The number density is $\rho=N/L^2=1.024$, where $L$ is the linear box length.

We perform Monte-Carlo (MC) dynamics
simulations composed of translational displacements
~\cite{frenkel2023understanding}. For the MC moves,
a particle is randomly chosen and displaced by a
vector randomly drawn within a square box of linear size
$\delta_{\rm max}=0.12$. The move is accepted according to the
Metropolis acceptance rule, satisfying the detailed balance condition.
Although MC simulations do not have an actual physical timescale, one can consider a time $t$ in units of MC sweeps, comprising $N$ MC trials. Indeed, it has been reported that Monte Carlo dynamics behaves essentially the same as other type of physical dynamics (e.g., Newtonian and Brownian dynamics) in terms of glassy slow dynamics~\cite{berthier2007monte}.
Thus, we analyze Monte Carlo dynamics following trajectories of particles and computing various time-dependent observables (see below).

To study dynamics of Monte-Carlo simulations, we measure the self-intermediate scattering function~\cite{kob1995testing}.
It is known that the dynamics of two-dimensional glass-forming liquids is strongly affected by long-wavelength fluctuations whose magnitude increases with the system size~\cite{shiba2016unveiling,illing2017mermin,vivek2017long,tarjus2017glass}. To remove this effect and study the intrinsic features of glassy dynamics, by convention, we employ the cage-relative version of the self-scattering function, which is
defined by
\begin{equation}
    F_{\rm s}^{\rm CR}(k,t) = \frac{1}{N} \sum_{i=1}^N \left\langle e^{-i {\bf k} \cdot \Delta {\bf r}^{\rm CR}_i(t)} \right\rangle_{\rm time} ,
    \label{label:Fskt}
\end{equation}
where $\langle \cdots \rangle_{\rm time}$ denotes the time average.
$\Delta {\bf r}^{\rm CR}_i(t)$ is the cage-relative displacement~\cite{shiba2016unveiling} given by 
\begin{equation}
\Delta {\bf r}^{\rm CR}_i(t)=\Delta {\bf r}_i(t)-\frac{1}{n_i} \sum_{j \in {NN}} \Delta {\bf r}_j(t) ,   
\end{equation}
where $\Delta {\bf r}_i(t)={\bf r}_i(t)-{\bf r}_i(0)$ is the displacement vector of the position ${\bf r}_i$ of the $i$-th particle and $n_i$ is the number of neighbor particles. The neighbors $NN$ are defined by the particles located within a circular cutoff whose radius is $1.4 \sigma_{\alpha \beta}$.
We note that the center of mass drifts at a longer time in the Monte-Carlo dynamics, and we subtracted such effect in our analysis.
We set $k=6.3$, which is close to the first peak of the static structure factor.

\begin{figure}
\centering
\includegraphics[width=0.98\columnwidth]{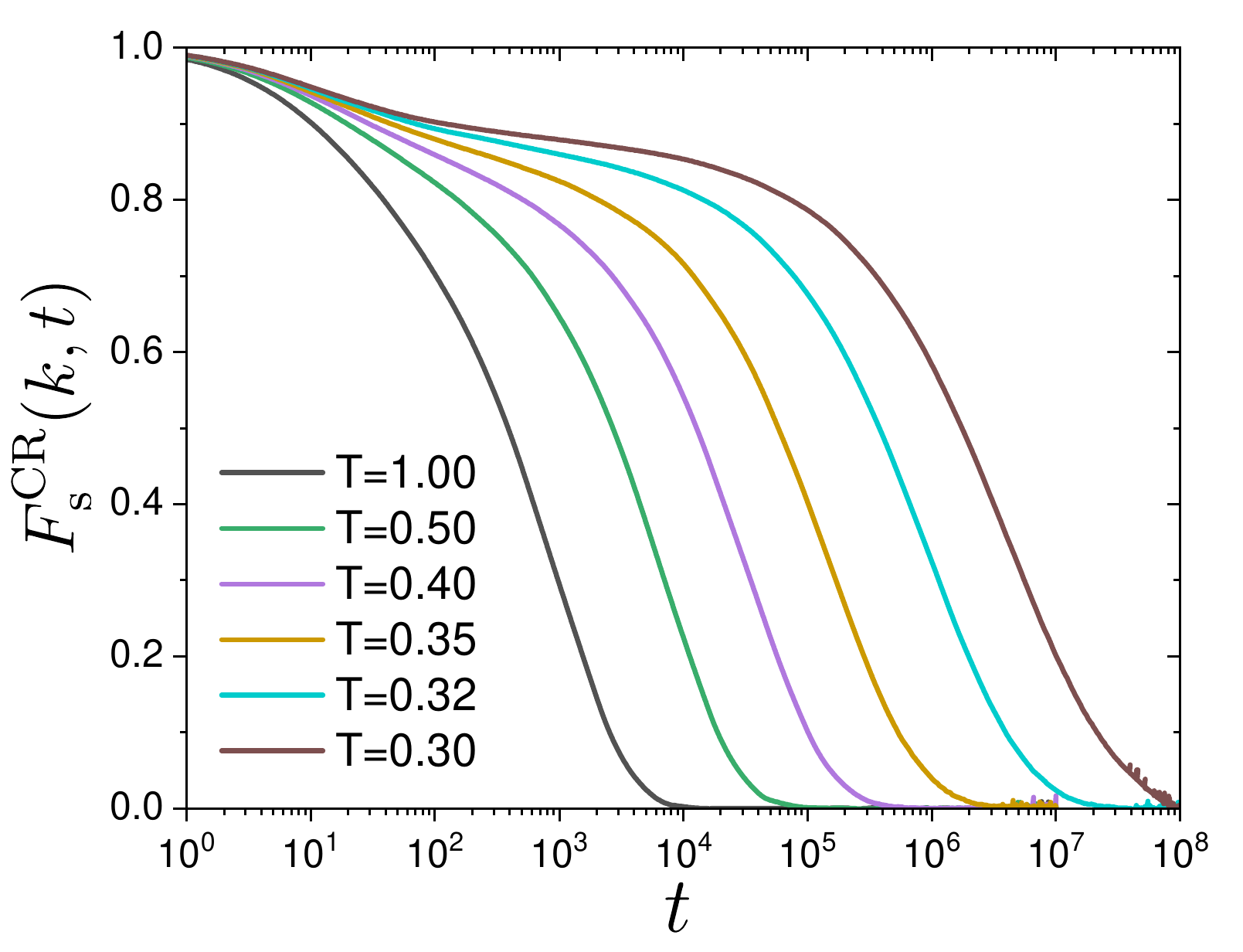}
\caption{Self-intermediate scattering function obtained by the Monte-Carlo dynamics.}
\label{fig:Fskt}
\end{figure}

We show $F_{\rm s}^{\rm CR}(k,t)$ in Fig.~\ref{fig:Fskt}.
$F_{\rm s}^{\rm CR}(k,t)$ relaxes slowly with decreasing temperature. In particular, it shows a two-step relaxation at lower temperatures, which is a hallmark of glassy slow dynamics observed in various other types of dynamics, such as Newtonian and Brownian dynamics~\cite{berthier2007monte}.
We then define a relaxation time $\tau_\alpha$ by $F_{\rm s}^{\rm CR}(k,\tau_\alpha)=1/e$.
$\tau_\alpha$ grows steeply, reflecting glassy slow dynamics, as expected.
In this study, we focus on $T=0.30$, which is the lowest temperature we can reach within our computational timescale.
The structural relaxation time $\tau_\alpha$ of this temperature is $\tau_\alpha \simeq 4 \times 10^6$.

\begin{figure}
\includegraphics[width=0.48\columnwidth]{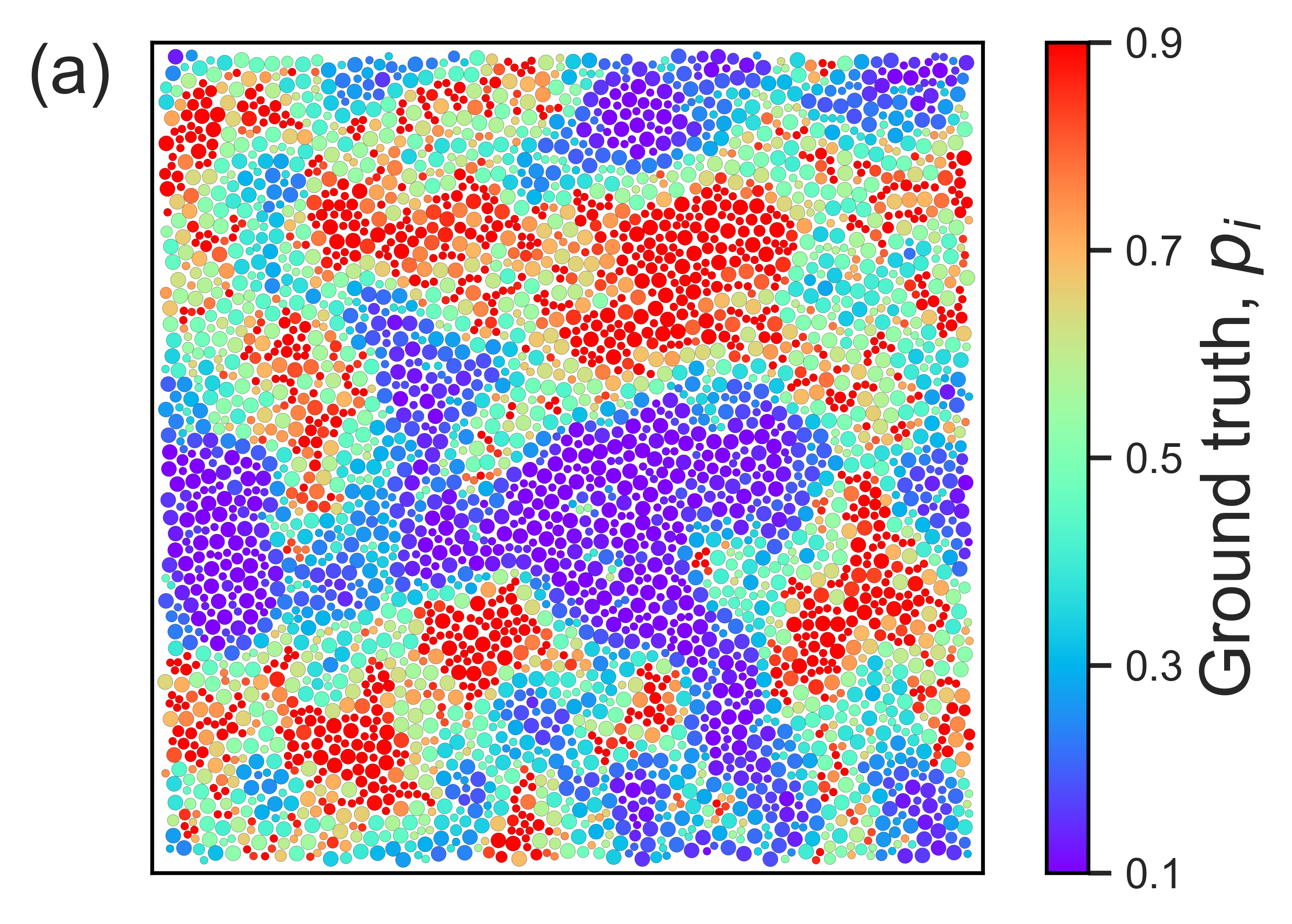}
\includegraphics[width=0.48\columnwidth]{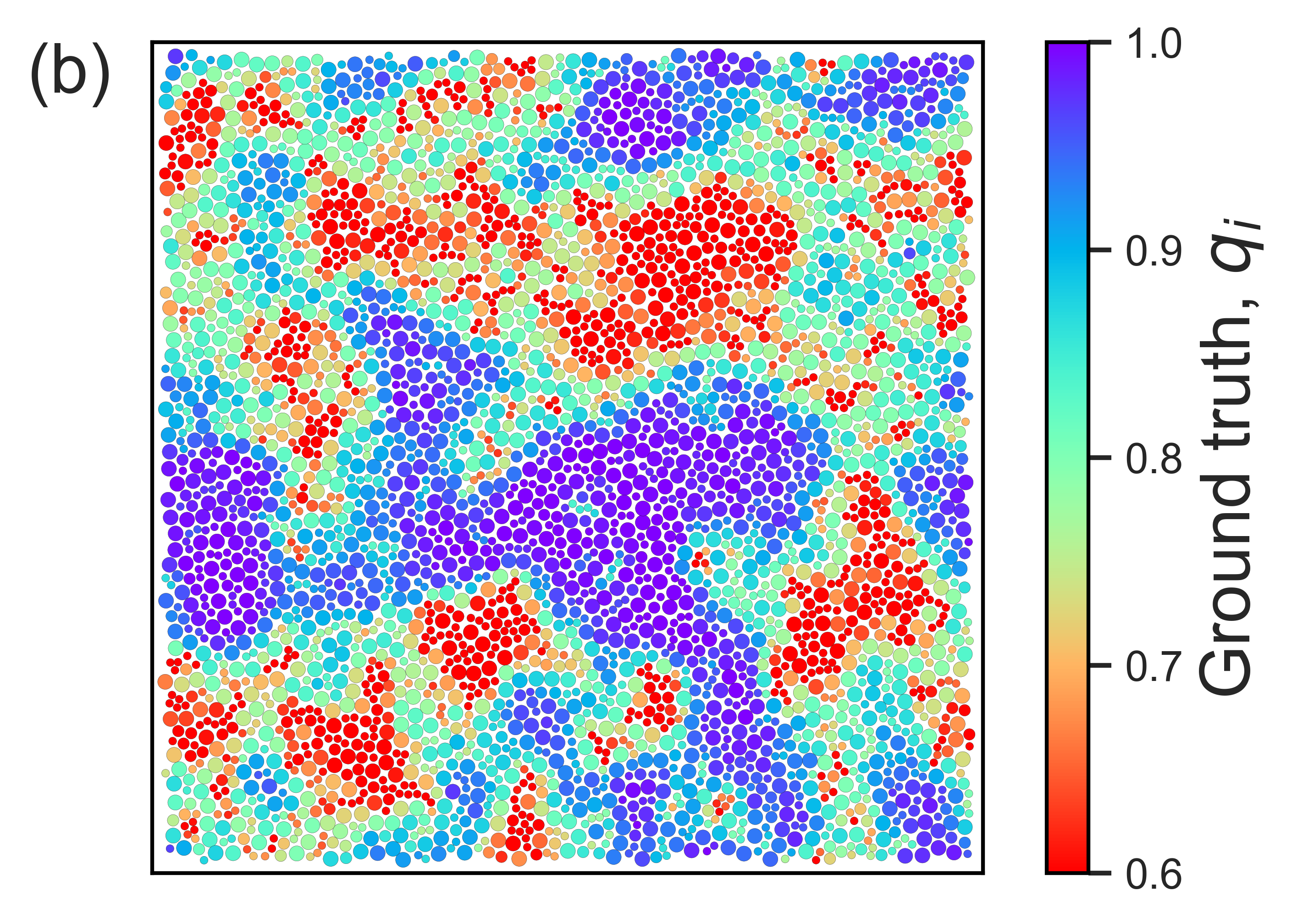}
\includegraphics[width=0.48\columnwidth]{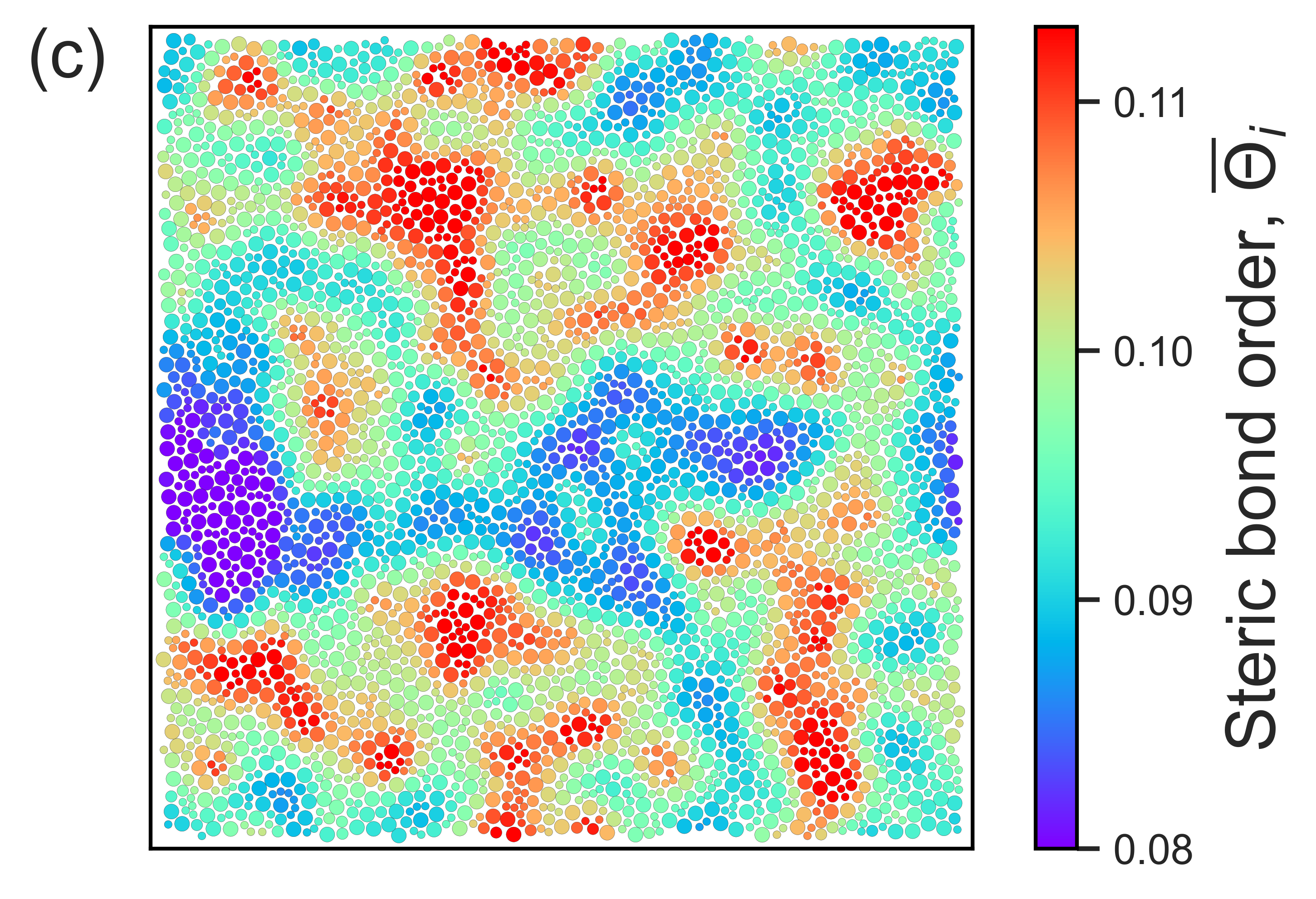}
\includegraphics[width=0.48\columnwidth]{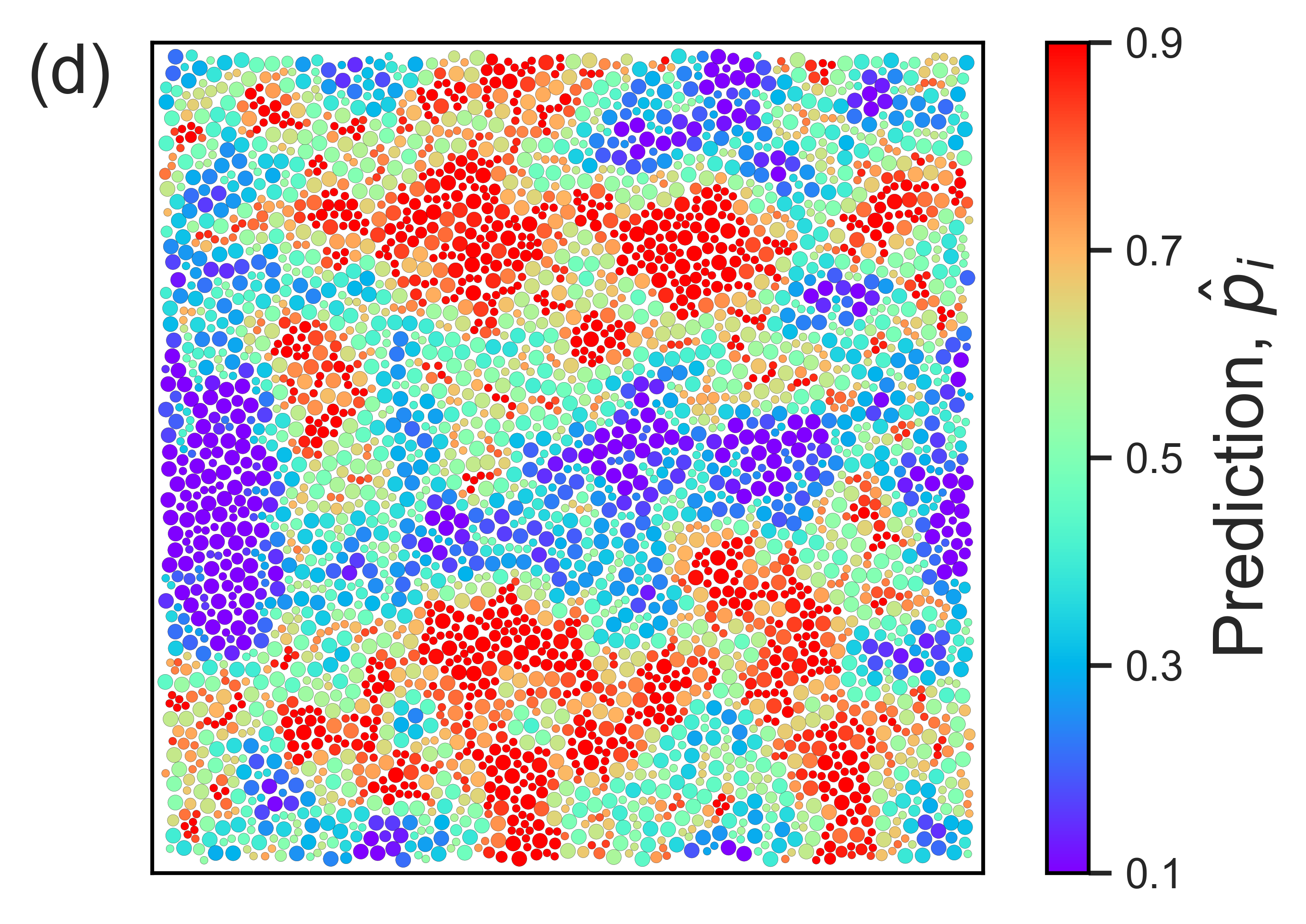}
\caption{(a, b): Map of dynamic propensities, $p_i(\tau_\alpha)$ measured by the displacement vectors (a) and $q_i(\tau_\alpha)$ measured by the bond-breaking (b). (c): Map of the steric bond order parameter, $\overline \Theta_i$, with the coarse-grained length $\ell=2.0$. (d): Map of the predicted propensity, $\hat p_i$, obtained by linear regression with $d=15$ features selected by the supervised information imbalance method in Sec.~\ref{sec:supervised}.} 
\label{fig:maps}
\end{figure}

To see heterogeneous dynamics in real space and assess the influence of static structure, we use the iso-configurational ensemble~\cite{widmer2006predicting,widmer2008irreversible}. We prepare an equilibrium configuration sampled at temperature $T=0.30$. We then perform the MC dynamics at $T=0.30$ many times with different seeds for random numbers, starting from the same initial equilibrium configuration. In particular, we compute a propensity parameter, $p_i$, which is given by
\begin{equation}
    p_i(t) = \left\langle |\Delta {\bf r}^{\rm CR}_i(t)| \right\rangle_{\rm iso}, 
\end{equation}
where $\langle \cdots \rangle_{\rm iso}$ is average over 30 different trajectories starting from a given static configuration.
In this paper, we mainly focus on the propensity of heterogeneous dynamics at the structural relaxation timescale, namely, $p_i(\tau_\alpha)$.
In Fig.~\ref{fig:maps}(a), we present $p_i(\tau_\alpha)$ showing dynamical heterogeneity.
Regions with larger and smaller values of $p_i$ correspond to mobile and immobile regions, respectively.

We also compute an another propensity parameter, $q_i(t)$, based on the bond-breaking approach~\cite{yamamoto1998dynamics,shiba2012relationship}.
 $q_i(t)$ is defined by
\begin{equation}
    q_i(t) = \left\langle n_i(t)/n_i(0) \right\rangle_{\rm iso} ,
    \label{eq:bond_breaking}
\end{equation}
where $n_i(0)$ is the
number of neighbor particles within a cutoff radius $1.4\sigma_{\alpha \beta}$
of the particle $i$ and $n_i(t)$ is the number of particles that were initially part of the neighbors (at $t=0$) and are still inside a cutoff $1.8\sigma_{\alpha \beta}$ at time $t$.
We present $q_i(\tau_\alpha)$ in Fig.~\ref{fig:maps}(b). 
$p_i(\tau_\alpha)$ and $q_i(\tau_\alpha)$ show very similar spatial patterns, suggesting that they share essentially the same information. 

\subsection{Physically motivated descriptors}

We consider various physically motivated descriptors characterizing the local structural environment.

The local potential energy $u_i$ for the particle $i$ is given by
\begin{equation}
u_i=\frac{1}{2} \sum_{j \neq i} v_{\alpha_i \beta_j}(r_{ij}) ,
\end{equation}
where $v_{\alpha_i \beta_j}(r_{ij})$ is the pair-wise Lennard-Jones potential which has a cutoff at $2.5 \sigma_{\alpha_i \beta_j}$.

The coordination number $z_i$ for the $i$-th particle is defined by the number of neighbor particles within $r_{ij}<1.5 \sigma_{\alpha_i \beta_j}$ (corresponding to near the first minimum of the radial distribution function).

We compute the conventional bond orientational order parameter in two-dimensions, $\Psi_{6, i}$, for the $i$-th particle, defined by
\begin{equation}
\Psi_{6, i} = \frac{1}{z_i} \left| \sum_{j=1}^{z_i} e^{\sqrt{-1}\ 6 \theta_{i j}} \right|,
\end{equation}
where $\theta_{ij}$ is the angle between ${\bf r}_{ij}={\bf r}_j-{\bf r}_i$ and the x-axis.
The condition for the nearest neighbors is again set to $r_{ij}<1.5 \sigma_{\alpha_i \beta_j}$.
$\Psi_{6, i}$ is often used to characterize hexagonal order in glassy and jamming systems~\cite{kawasaki2007correlation,schreck2011tuning}.
$\Psi_{6, i}$ takes $1$ for perfectly hexagonal packings, where as $\Psi_{6, i}$ takes lower values for disordered packings.

We also compute a steric bond order parameter $\Theta_i$ introduced by Tong and Tanaka~\cite{tong2018revealing}.
Here we consider a central particle $i$ and its neighbor particles.
The definition of the neighbors is the same as $\Psi_{6, i}$.
For each pair $\langle jk \rangle$ of neighbor particles next to each
other, we measure the angle between ${\bf r}_{ij}$ and ${\bf r}_{ik}$, denoted as
$\theta_{jk}$.  
The reference configuration with these three particles, $i$, $j$, and $k$, perfectly
just in touch, produces the angle
indicated as $\theta_{jk}^{\rm ref}$. 
Practically $\theta_{jk}^{\rm ref}$ is computed by $\sigma_{\alpha_i \beta_j}$, $\sigma_{\alpha_i \beta_k}$, and $\sigma_{\alpha_j \beta_k}$, using the cosine formula.
Then we define the order parameter for the $i$th
particle as
\begin{equation}
\Theta_i = \frac{1}{z_i} \sum_{\langle jk \rangle} \left| \theta_{jk}-\theta_{jk}^{\rm ref} \right| ,
\end{equation}
where $\langle jk \rangle$ denotes the summation over all pairs of neighbors.
If particles form stericallly favored, well-packed configurations, $\Theta_i$ produces smaller value, because $\theta_{jk}$ would be close to the reference, $\theta_{jk}^{\rm ref}$.
Instead, disordered packings generally take larger $\Theta_i$, since $\theta_{jk}$ would strongly deviate from $\theta_{jk}^{\rm ref}$.
Thus, $\Theta_i$ characterizes amount of disorder, which would play the similar roles as $1-\Psi_{6, i}$, yet $\Theta_i$ is more sensitive order parameter for multi-components (or polydisperse) systems~\cite{tong2018revealing}.

We next consider the coarse-grained local number density,
\begin{equation}
    \overline \rho_i(\ell) = \sum_{j \in \mathcal{N}_i} e^{-r_{ij}/\ell} ,
\end{equation}
where $\mathcal{N}_i$ is the set of particles of $i$ including $i$ itself and $\ell$ is the coarse-graining lengthscale. These particles are inside the circle with the radius $R_\mathcal{N}$ centered at the position of $i$. We set $R_\mathcal{N}=20 \sigma_{\rm LS}$ in this study.

We also compute local volume (area) fraction,
\begin{equation}
    \overline \varphi_i(\ell) = \sum_{j \in \mathcal{N}_i} (\sigma_{\alpha_i \beta_j})^2 e^{-r_{ij}/\ell} .
\end{equation}

We perform the above coarse-graining procedure for all the other descriptors, $x_i=u_i, \ z_i, \ \Psi_{6, i}$, and $\Theta_i$, as given by
\begin{equation}
    \overline x_i(\ell) = \left( \overline \rho_i \right)^{-1} \sum_{j \in \mathcal{N}_i} x_j e^{-r_{ij}/\ell}  ,
\end{equation}
which provides us with $\overline u_i(\ell)$, $\overline z_i(\ell)$, $\overline \Psi_{6, i}(\ell)$, and $\overline \Theta_i(\ell)$.
We vary the coarse-graining lengthscale $\ell$ from $\ell=0.5$ to $5.0$ for each descriptor.
Thus in total, we have $M=60$ features (6 different physically motivated descriptors times 10 different coarse-grainning lengthscales).

For each feature, we normalize it to have zero mean and unit variance. After normalization, all features constitute a feature vector ${\bm X}_{{\rm Full}, i}$ for a particle $i$, given by
\begin{equation}
    {\bm X}_{{\rm Full}, i}=\left(X_i^{(1)}, \ X_i^{(2)}, \ \cdots, \ X_i^{(M)}\right) .
\end{equation}
This is the structural input for the information imbalance analysis described below.

\subsection{Information imbalance}

\begin{figure*}
\includegraphics[width=0.7\columnwidth]{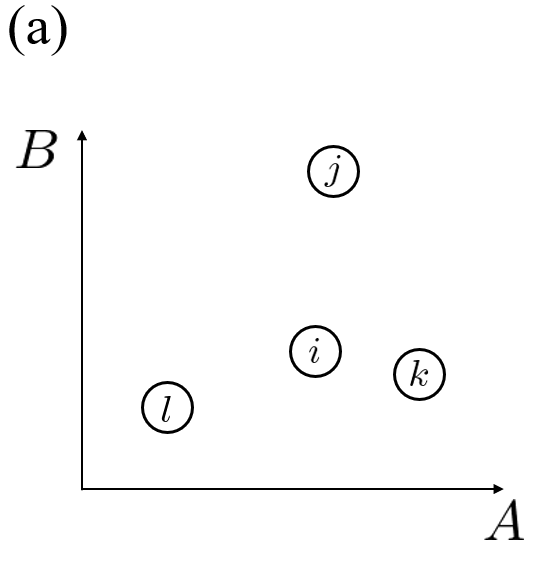}
\includegraphics[width=0.75\columnwidth]{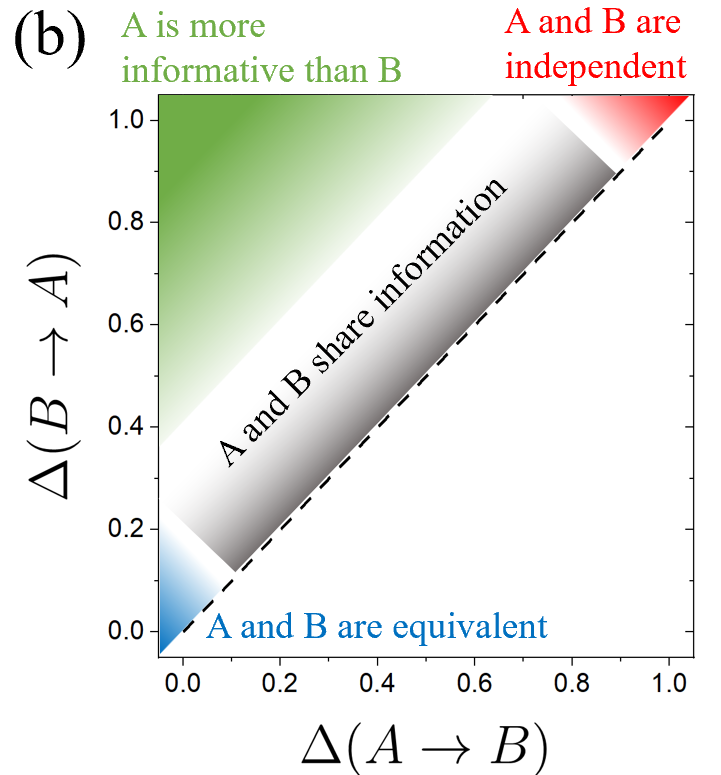}
\caption{(a): Schematic illustration of data points in the feature space composed of feature $A$ and $B$. (b): Information imbalance plane. Four distinct relationships between $A$ and $B$ are presented schematically (see the main text).  }
\label{fig:concept_information_imbalance}
\end{figure*}

We recap the information imbalance introduced in Ref.~\onlinecite{glielmo2022ranking}.
Suppose that we have two features (physically motivated descriptors in our case), $A$ and $B$, characterizing a data point.
We wish to ask if $A$ and $B$ are equivalent, independent, or if one is more informative than the other.
The last question, i.e., $A$ is more informative than $B$, is particularly challenging for standard statistical analysis where $A$ and $B$ are treated equally or symmetrically. Recall, for example, that the Pearson coefficient $\rho_{A, \ B}$ (see Sec.~\ref{sec:Pearson_vs_IB} for the definition) and mutual information $I(A,\ B)$ are defined in a {\it symmetric} way, $\rho_{A, \ B}=\rho_{B, \ A}$ and $I(A,\ B)=I(B, \ A)$, respectively. Hence, although $\rho_{A,\ B}$ and $I(A,\ B)$ can quantify the correlation between $A$ and $B$, they alone cannot tell, say, $A$ is more informative than $B$.

The information imbalance analysis between $A$ and $B$ is constructed in an {\it asymmetric} way, which allows us to tell if one is more informative than the other.
We first consider a feature space spanned by $A$ and $B$ as schematically shown in Fig.~\ref{fig:concept_information_imbalance}(a).
Each of $A$ and $B$ can generally be vectors, but for simplicity, we consider both to be scalars, forming a two-dimensional plane.
Here we have four data points labeled by $i$, $j$, $k$, and $l$, respectively.
Let us focus on the data point $i$ and make a ranking based on distance along the $A$-axis and the $B$-axis separately.
We define the Euclidean distance by $d_{ij}^A=|A_i-A_j|$ and $d_{ij}^B=|B_i-B_j|$, where $A_i$ and $B_i$ are the $A$ and $B$ component of the data point $i$. Note that one can also use non-Euclidean distance in general.
For example, in Fig.~\ref{fig:concept_information_imbalance}(a), data point $j$ is the closest data point of $i$ in terms of $A$, which is ranked first and denoted as $r_{ij}^A=1$.
In general, $r_{ij}^A$ is the rank of $j$ relative to $i$, obtained by sorting the pairwise distances between $i$ and rest of the points from
smallest to largest. The data points $k$ and $l$ are the second and third closest data points in terms of $A$, which are written as $r_{ik}^A=2$ and $r_{il}^A=3$, respectively.
Similary, one can construct ranking based solely on $B$, which are written as $r_{ij}^B=3$, $r_{ik}^B=1$, and $r_{il}^B=2$.

We now imagine a situation where $A$ contains information about $B$ (to some extent). In such a case, the organization of $N$ data points in the feature space would be as follows. When two data points, say $i$ and $j$, are close along $A$, they would also be close along $B$.
This idea can be quantified by monitoring an average rank of $r_{ij}^B$ when $r_{ij}^A=1$, i.e., $\frac{1}{N} \sum_{i=1}^N r_{ij}^B$ given $r_{ij}^A=1$. If $r_{ij}^B$ is close to $1$ for most of $i$, we would get $\frac{1}{N} \sum_{i=1}^N r_{ij}^B \simeq 1$ on average.
Instead, when $A$ contains no information about $B$, $r_{ij}^B$ (given $r_{ij}^A=1$) would be completely random, and hence we get $\frac{1}{N} \sum_{i=1}^N r_{ij}^B \simeq N/2$. We will use $N/2$ for a normalization factor below.

Finally, based on the above consideration, we define the information imbalance $\Delta(A \to B)$ using the average rank of $r_{ij}^B$ given $r_{ij}^A=1$ with the normalization factor:
\begin{eqnarray}
\Delta(A \to B) &=& \left( \frac{N}{2} \right)^{-1}\frac{1}{N} \sum_{i=1}^N \sum^N_{\substack{j=1 \\ ({\rm s.t.} \ r_{ij}^A=1 )}} r_{ij}^B	\nonumber \\ 
 &=& \frac{2}{N^2} \sum^N_{i=1} \sum^N_{j=1} \delta_{1, r_{ij}^A} \ r_{ij}^B  ,
\label{eq:def_information_imbalance}
\end{eqnarray}
where $\delta_{1, r_{ij}^A}$ is the Kronecker's delta selecting $r_{ij}^A=1$. We also used $r_{ii}^A=0$ and $r_{ii}^B=0$.
Because of the normalization factor, $N/2$, we get $\Delta(A \to B) \simeq 0$ when $A$ predicts the ranking of $B$ strongly, namely, $A$ contains information about $B$, whereas $\Delta(A \to B) \simeq 1$ when $A$ cannot determine the ranking of $B$ at all, or $A$ is not informative about $B$.

As one expect from the definition in Eq.~(\ref{eq:def_information_imbalance}),
one can also define $\Delta (B \to A)$ by computing the average rank of $r_{ij}^A$ given $r_{ij}^B=1$.
By measuring both $\Delta (A \to B)$ and $\Delta (B \to A)$ together, we can finally ask if $A$ and $B$ are equivalent, independent, or if one is more informative than the other.
The limiting cases relevant to our study is summarized as follows~\cite{glielmo2022ranking}.
\begin{itemize}
    \item {\bf $A$ and $B$ are equivalent, carrying identical information when $\Delta(A \to B) \simeq 0$ and $\Delta(B \to A) \simeq 0$.}
    \item {\bf $A$ and $B$ are independent (or orthogonal), carrying completely independent information when $\Delta(A \to B) \simeq 1$ and $\Delta(B \to A) \simeq 1$.}
    \item {\bf $A$ is more informative than $B$ when $\Delta(A \to B) \simeq 0$ and $\Delta(B \to A) \simeq 1$.}
    \item {\bf $A$ and $B$ share information symmetrically, carrying both identical and independent information when $0 \ll \Delta(A \to B) = \Delta(B \to A) \ll 1$.} 
\end{itemize}
In practice, it is convenient to plot a two-dimensional plane of $\Delta(A \to B)$ and $\Delta(B \to A)$, called the information imbalance plane, as shown in Fig.~\ref{fig:concept_information_imbalance}(b). 

Up to here, we considered information imbalance between $A$ and $B$. In applications, one often studies information imbalance between a feature (vector) composed of $A$ and $B$, denoted as $(A,B)$, and a (scalar) feature $C$. In such a case, one has to define the distances properly. For example, one can define distances for the features, $(A,B)$ and $C$, by
 $d_{ij}^{(A, B)}=\sqrt{(A_i-A_j)^2 + (B_i-B_j)^2}$ and $d_{ij}^C=|C_i-C_j|$, respectively, using the Euclidean distance.
Then the ranking $r_{ij}^{(A,B)}$ and $r_{ij}^{C}$ can be obtained based on $d_{ij}^{(A, B)}$ and $d_{ij}^{C}$, respectively. The above setting allows us to compute $\Delta\left((A,B) \to C\right)$ and $\Delta\left(C \to (A,B)\right)$ appropriately.

In this paper, we compute the information imbalance, $\Delta (A \to B)$ and $\Delta (B \to A)$, as well as the Pearson coefficient (see below), for each configuration composed of $N=4000$ particles. We average them over 10 different configurations.

\section{Results}
\label{sec:Results}

\subsection{Pearson coefficient and information imbalance}
\label{sec:Pearson_vs_IB}

We first analyze the relationship between dynamic propensity and physically motivated descriptors by using the conventional Pearson coefficient $\rho_{A, \ B}$, as a reference for the information imbalance analysis below.
$\rho_{A, \ B}$ quantifies the correlation between two variables $A$ and $B$:
\begin{equation}
    \rho_{A, \ B} = \frac{1}{N} \sum_{i=1}^{N} \frac{(A_i - \mu_A)(B_i - \mu_B)}{\sigma_A \sigma_B} ,
\end{equation}
where $\mu_A$ and $\mu_B$ are the mean of $A_i$ and $B_i$, respectively, and $\sigma_A$ and $\sigma_B$ are the standard deviation of $A_i$ and $B_i$, respectively.

\begin{figure*}
\includegraphics[width=0.8\columnwidth]{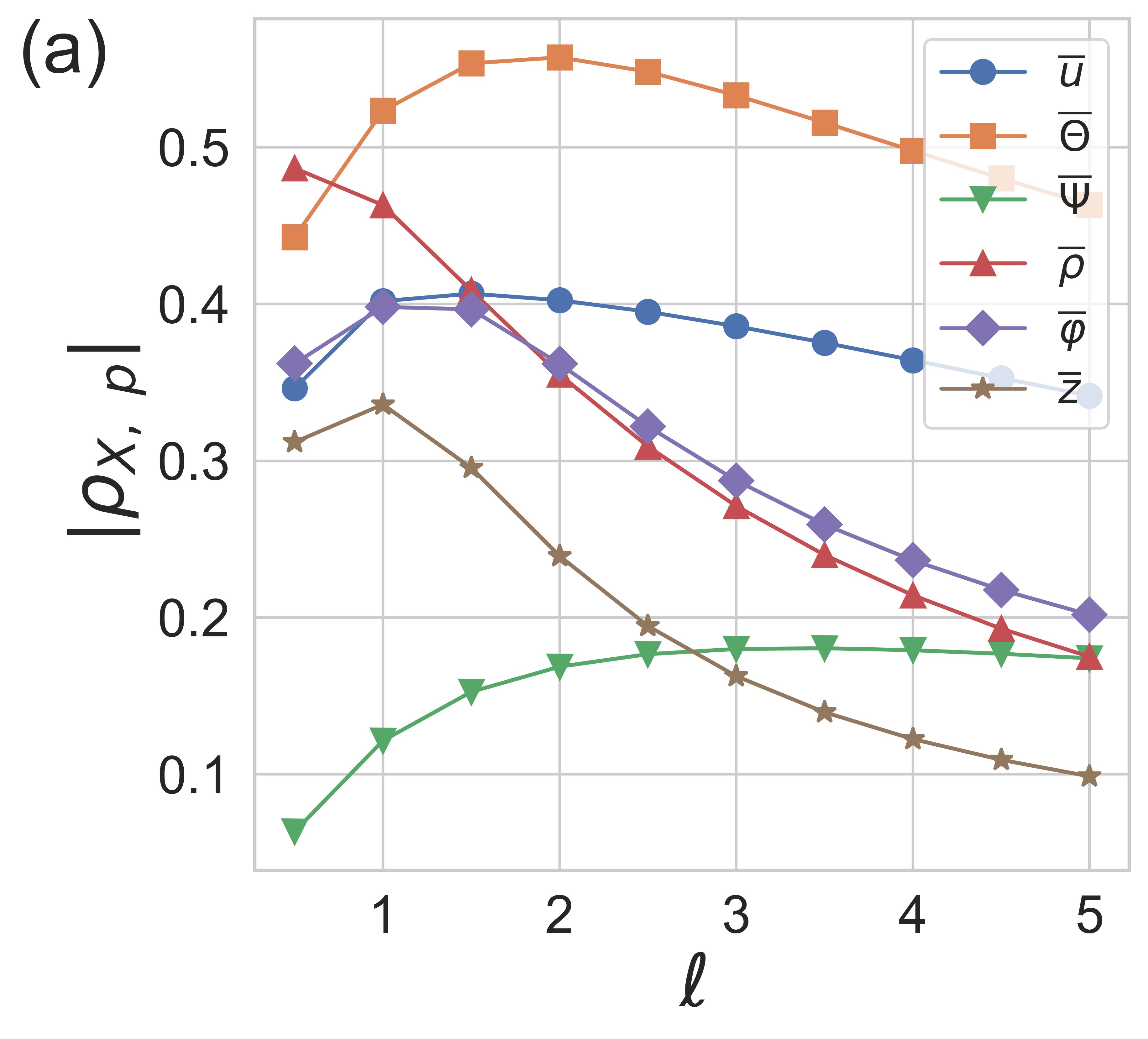}
\includegraphics[width=0.8\columnwidth]{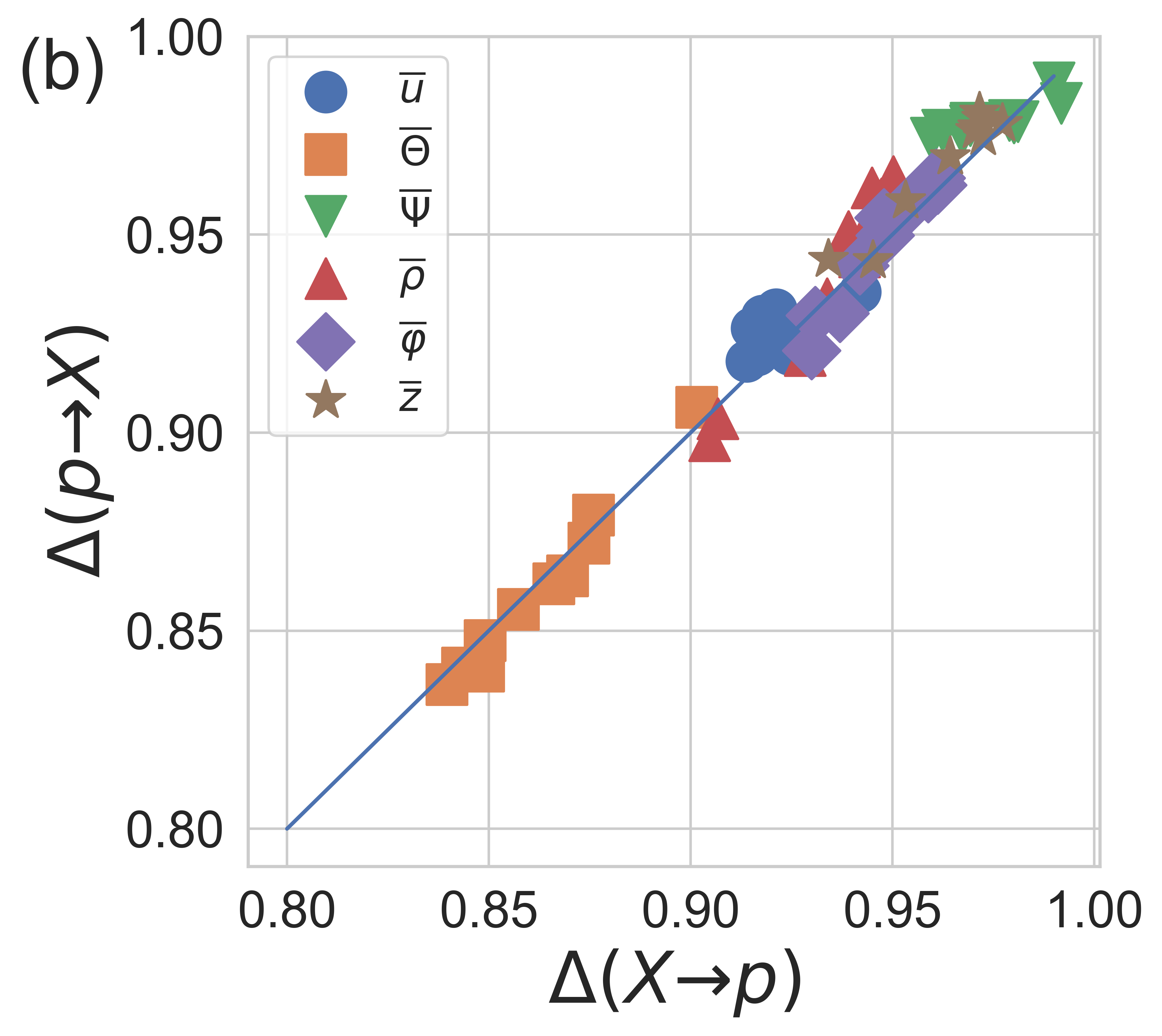}
\includegraphics[width=0.8\columnwidth]{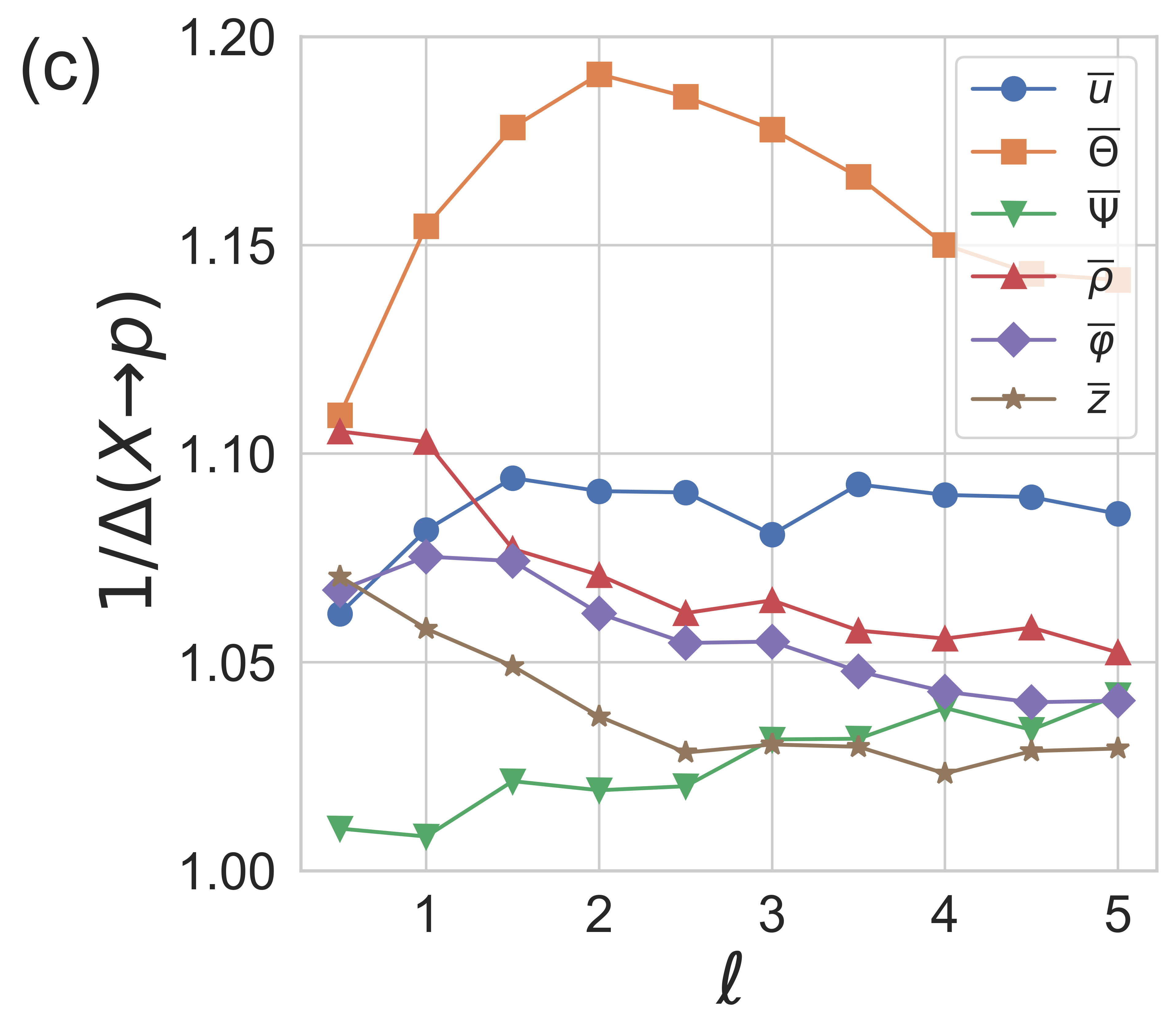}
\includegraphics[width=0.8\columnwidth]{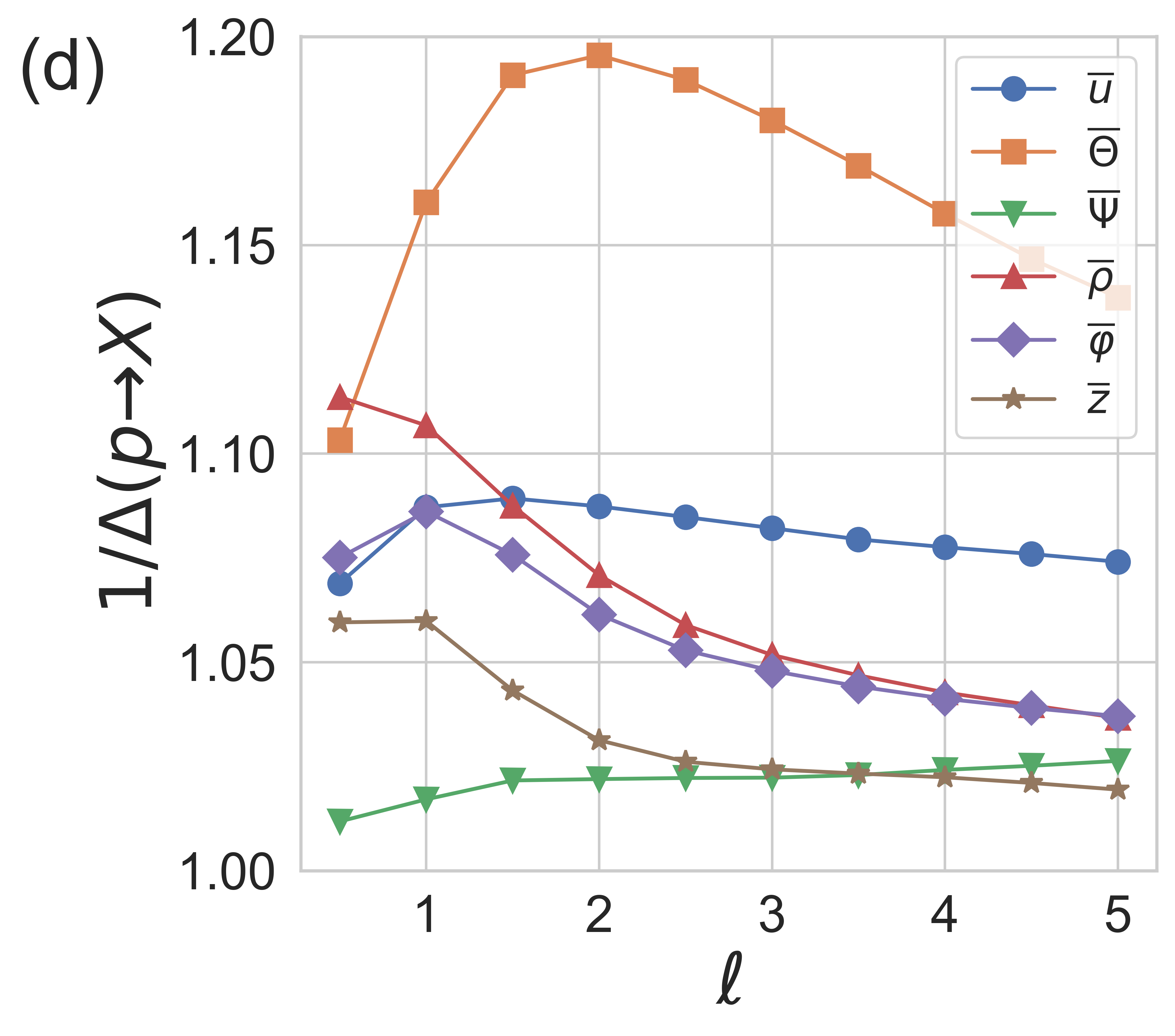}
\caption{(a): The absolute value of the Pearson coefficient between the propensity $p_i$ and various physical descriptors as a function of the coarse-grained length $\ell$. (b): Information imbalance plane composed of $\Delta (X \to p)$ and $\Delta (p \to X)$. Each physically motivated descriptor has 10 data points associated with different coarse-graining lengths (shown in the same color). (c, d): The inverse of the information imbalance $\Delta (X \to p)$ (c) and $\Delta (p \to X)$ (d).}
\label{fig:pearson_vs_IB}
\end{figure*}

Figure~\ref{fig:pearson_vs_IB}(a) shows the Pearson coefficient, $\rho_{X, \ p}$, between the propensity $p_i(\tau_\alpha)$ and all descriptors  we considered (here simply denoted as $X$) with varying the coarse-graining lengthscale $\ell$.
Note that because the volume fraction and coordination number have anti-correlation with propensity, we present the absolute value of the Pearson coefficient. 
A larger value of $|\rho_{X, \ p}|$ corresponds to a higher correlation (or anti-correlation)  between the propensity and descriptor $X$.

We find that $\overline \Theta$ is the best descriptor in the plot, showing a non-monotonic $\ell$ dependence. $\rho_{\overline \Theta, \ p}$ has the maximum value (around $0.56$) at an optimal lengthscale near $\ell = 2$, consistent with previous literature~\cite{tong2018revealing}. We show in Fig.~\ref{fig:maps}(c) a map for $\overline \Theta_i(\ell=2)$. Indeed, it shows a similar pattern as the dynamical heterogeneity presented in Fig.~\ref{fig:maps}(a).
The other descriptors demonstrate a smaller correlation with dynamics either monotonic ($\overline \rho$, $\overline \Psi$) or non-monotonic ($\overline u$, $\overline \varphi$, $\overline z$) dependence on $\ell$.

Next, we analyze the same dataset by the information imbalance and ask if we get the same trend as the Pearson coefficient.
Recall that when $\Delta(X \to p)$ is small, a feature $X$ predicts the propensity $p$ well. 
Hence, to compare with the Pearson coefficient, we plot the inverse of $\Delta(X \to p)$ in Fig.~\ref{fig:pearson_vs_IB}(c).
We find that $1/\Delta(X \to p)$ follows a quite similar trend as the Pearson coefficient in terms of the order of the descriptors as well as their $\ell$ dependence.

We also measure $\Delta(p \to X)$ and plot its inverse in Figs.~\ref{fig:pearson_vs_IB}(d). Interestingly, $1/\Delta(p \to X)$ also shows essentially the same behavior with a similar value, suggesting $\Delta(X \to p) \simeq \Delta(p \to X)$.
Thus, in Fig.~\ref{fig:pearson_vs_IB}(b), we present the $\Delta(X \to p)$ versus $\Delta(p \to X)$ plot (information imbalance plane) for all features studied. 
Indeed, we find that all features concentrate on the straight line having a symmetrical form, $\Delta(X \to p) \simeq \Delta(p \to X)$.
In particular, $\overline \Psi_6$ (for all $\ell$) takes $\Delta(\overline \Psi_6 \to p) \simeq \Delta(p \to \overline \Psi_6) \simeq 1$, meaning that $\overline \Psi_6$ and propensity $p$ are independent (or orthogonal), sharing no information, at least in the simulation model we use.
Instead, $\overline \Theta$ shows 
$\Delta(\overline \Theta \to p) \simeq \Delta(p \to \overline \Theta) \simeq 0.85$, meaning that $\overline \Theta$ and propensity $p$ share information to some extent.
These trends are consistent with findings in the Pearson coefficient.

To better understand the information imbalance plane, we also compute the information imbalance between $p$ and the propensity $q$ measured by the bond-breaking in Eq.~(\ref{eq:bond_breaking}). 
We obtain $\Delta(q \to p) \simeq 0.21$ and $\Delta(p \to q) \simeq 0.26$, meaning that $p$ and $q$ share very similar information, as expected from the visual comparison between Figs.~\ref{fig:maps}(a) and (b).

To conclude, the information imbalance analysis can characterize the relationship between structural features and dynamic propensity, reproducing the trend obtained by the Pearson coefficient. 
We also found that the information imbalance between each feature and propensity is symmetric.
This means that although some descriptors such as $\overline \Theta$ and $\overline u$ have a higher correlation with propensity, no {\it single} feature is more informative than the propensity. 
This is the first non-trivial statement achieved by the information imbalance analysis.

\subsection{Feature selection by supervised learning}

\label{sec:supervised}

We next perform feature selection among a pool of $M=60$ physically motivated descriptors by information imbalance in a supervised learning manner, using propensity data.
This method is inspired by feature selection of relevant policies on the Covid-19 epidemic state in terms of the number of deaths in the future~\cite{glielmo2022ranking}.
Recall that the full set of $M$ features is denoted as the vector, ${\bm X}_{\rm Full}$, which is given by
\begin{equation}
    {\bm X}_{\rm Full} = (X^{(1)}, \ X^{(2)}, \ \cdots, \ X^{(M)}) .
\end{equation}
Here, we drop off the particle index $i$ for simplicity.
The goal is to select $d \ll M$ features out of $M=60$ features and obtain a vector of selected features,
\begin{equation}
    {\bm X}_d = (X^{\rm 1st}, \ X^{\rm 2nd}, \ \cdots, \ X^{d{\rm th}}) .
    \label{eq:selected_features}
\end{equation}
This operation corresponds to a dimensional reduction from $M$ to $d$.

The algorithm to obtain ${\bm X}_d$ using the information imbalance is as follows.
First, among $M(=60)$ descriptors, we search $X^{\rm 1st}$ at which the information imbalance between the feature and propensity is minimized at $\Delta ({\bm X}_1 \to p)$, where ${\bm X}_1=(X^{\rm 1st})$.
We next search $X^{\rm 2nd}$ among the rest of $M-1$ features such that the imformation imbalance is minimized at $\Delta ({\bm X}_2 \to p)$, where ${\bm X}_2=(X^{\rm 1st}, \ X^{\rm 2nd})$.
We repeat the above procedure until we find $X^{d{\rm th} }$ among the rest of $M-(d-1)$ features such that the information imbalance is minimized at $\Delta ({\bm X}_d \to p)$. For each minimization, we perform a greedy algorithm to find the optimal feature.
Since ${\bm X}_d$ is found to have minimum at $\Delta ({\bm X}_d \to p)$, at least within our algorithm, ${\bm X}_d$ should have a high predictivity in terms of the propensity $p$.
Note that in the above iterative procedure, we keep the previously selected features and find a new one among the rest of the features. 
Although it is not guaranteed to reach the global minimum of $\Delta ({\bm X}_d \to p)$,
this procedure avoids searching all combinatorially possible sets of $d$ features, which requires a huge computational cost~\cite{glielmo2022ranking}.

\begin{figure}
\includegraphics[width=0.98\columnwidth]{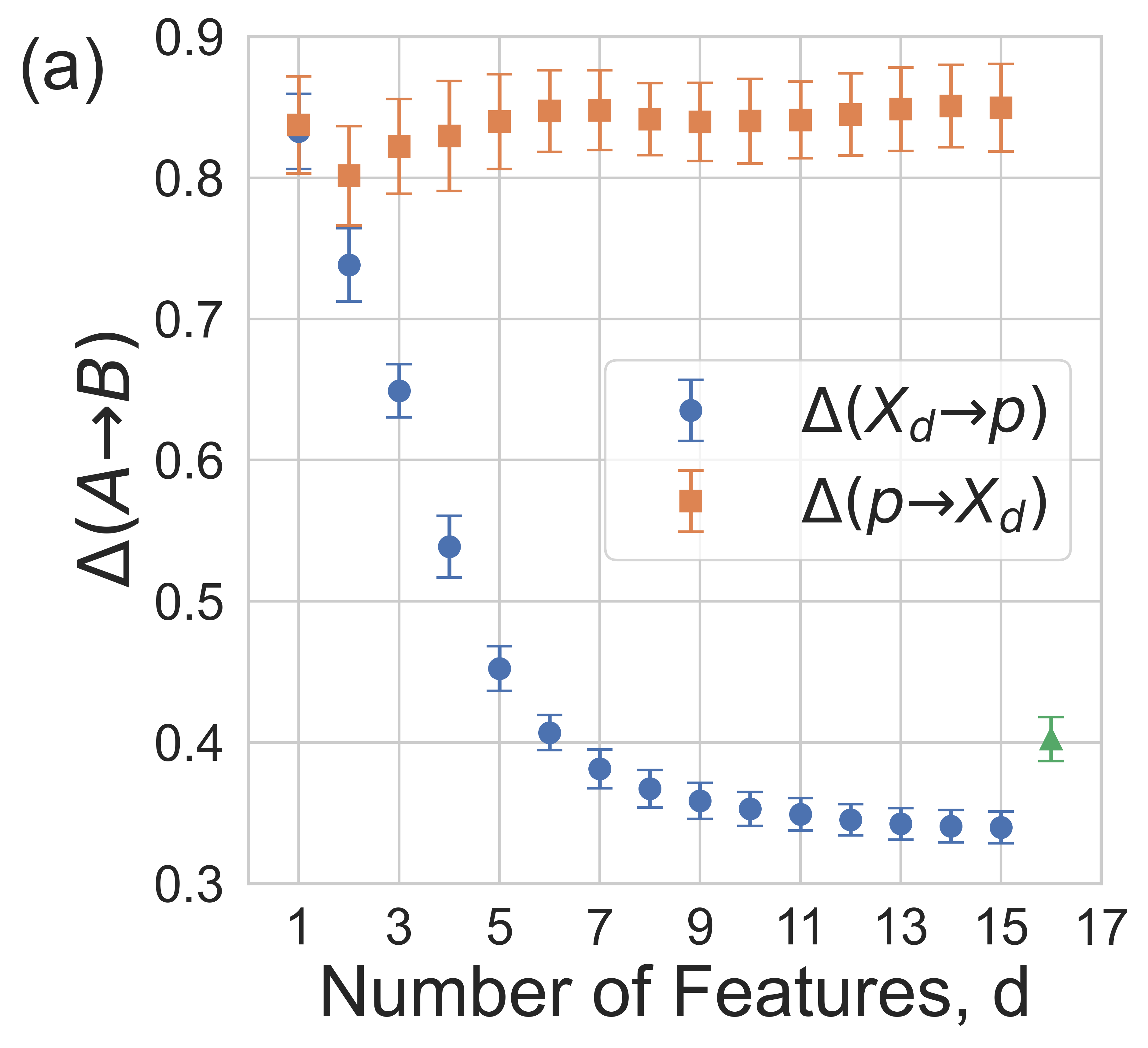}
\includegraphics[width=0.98\columnwidth]{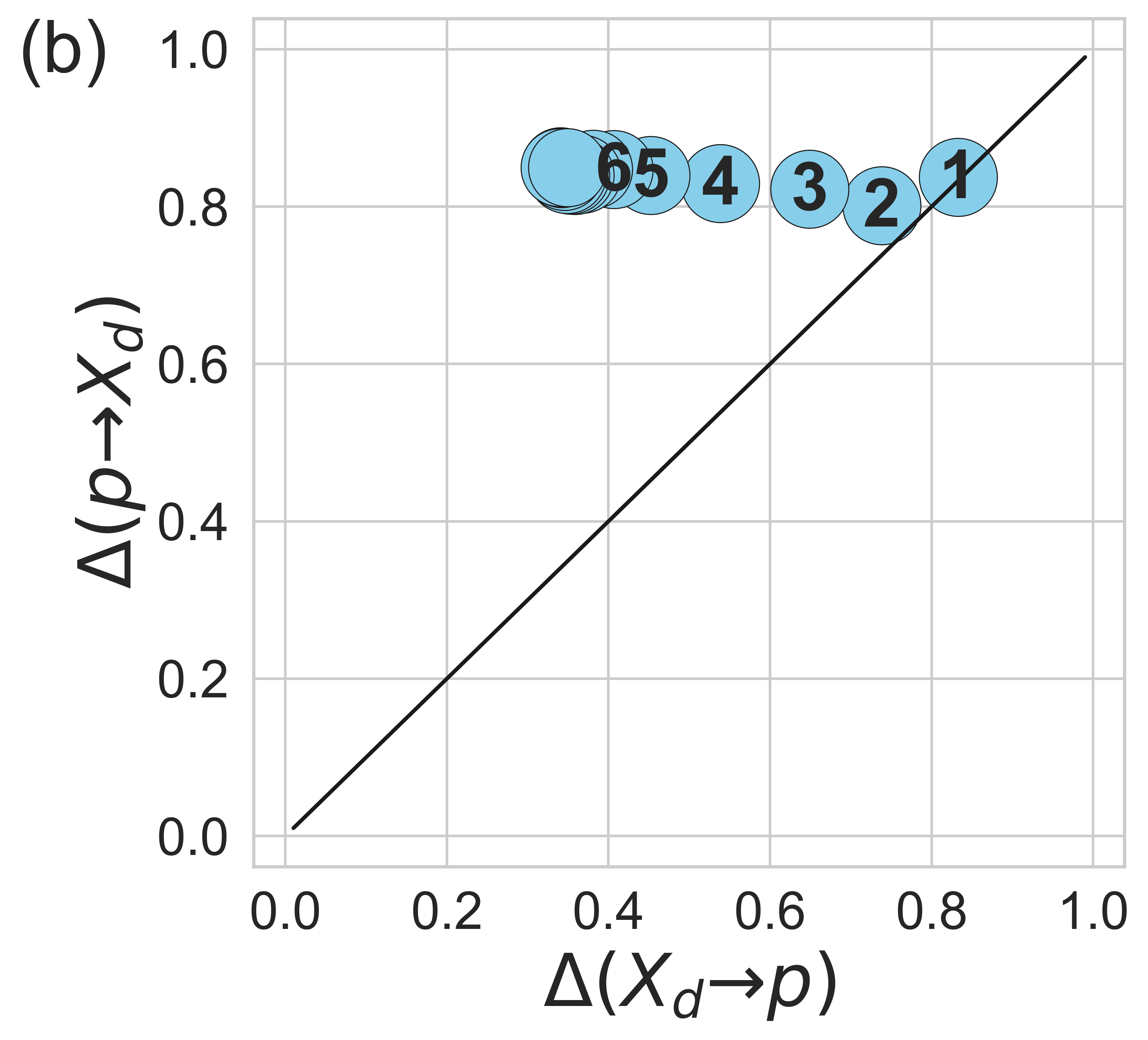}
\caption{Feature selection by the supervised approach. (a): Information imbalance, $\Delta ({\bm X}_d \to p)$ and $\Delta (p \to {\bm X}_d)$, as a function of the number of selected features, $d$. The errorbars correspond to the standard deviation from 10 different configurations. The last triangular data point at $d=16$ represents the addition of one feature generated from Gaussian noise. (b): The same data are presented in the information imbalance plane. The number inside a point represent $d$.} 
\label{fig:IB_supervised}
\end{figure}

Figure~\ref{fig:IB_supervised}(a) shows $\Delta ({\bm X}_d \to p)$ as a function of $d$. $\Delta ({\bm X}_d \to p)$ decays quickly as $d$ is increased and it saturates, say, above $d=10$. This means that ${\bm X}_{d=10}$ is already as informative as the full set, ${\bm X}_{d=M}={\bm X}_{\rm Full}$, in terms of prediction of the propensity.
The selected features up to $d=10$ from the 1st to $d$th, for the configuration presented in Fig.~\ref{fig:maps} for instance, are as follows: $\overline{\Theta_i}(\ell=1.5)$, $\overline{\rho_i}(\ell=1)$, $\overline{\rho_i}(\ell=4)$, $\overline{\Theta_i}(\ell=5)$, $\overline{u_i}(\ell=5)$, $\overline{\varphi_i}(\ell=4.5)$, $\overline{z_i}(\ell=1.5)$, $\overline{\Psi}_{6, i}(\ell =4)$, $\overline{\varphi_i}(\ell = 0.5)$, and  
$\overline{z_i}(\ell=5)$.
This lineup fluctuates slightly depending on configurations, yet the overall trend does not change.

One might assume that adding features, even irrelevant ones, would decrease (or at least not increase) $\Delta(A \to B)$. However, this is not always the case, as demonstrated in the following exercise~\cite{glielmo2022ranking}. In $\Delta ({\bm X}_d \to p)$, we add an independent feature generated from Gaussian noise, which is entirely irrelevant. As shown in Fig.~\ref{fig:IB_supervised}(a), $\Delta ({\bm X}_d \to p)$ actually increases with the addition of this irrelevant feature.

Instead, as seen in Fig.~\ref{fig:IB_supervised}(a), $\Delta (p \to {\bm X}_d)$ is essentially flat as a function of
$d$, which means that there is no improvement of predictivity from $p$ to ${\bm X}_d$, as expected.
This huge asymmetry can be seen in the information imbalance plane in Fig.~\ref{fig:IB_supervised}(b), where we show the evolution of $\left(\Delta ({\bm X}_d \to p), \ \Delta (p \to {\bm X}_d) \right)$ with increasing $d$.
The first data point at $d=1$ associated with a single feature ${\bm X}_1=(X^{\rm 1st})$ locates on the straight line, showing symmetry, $\Delta ({\bm X}_1 \to p) \simeq \ \Delta (p \to {\bm X}_1)$, as we have already seen in Sec.~\ref{sec:Pearson_vs_IB}.
Yet, with increasing the number of selected features, $\left(\Delta ({\bm X}_d \to p), \ \Delta (p \to {\bm X}_d) \right)$ deviates from the straight line with a strong asymmetry, $\Delta ({\bm X}_d \to p) \ll \Delta (p \to {\bm X}_d)$. This means that ${\bm X}_d$ is getting more and more informative than propensity $p$ with increasing $d$, while $p$ alone is getting harder to predict ${\bm X}_d$.
From this observation, we conclude that {\it multiple} features are required to construct a set of selected features that is more informative than the dynamic propensity.

The above trend is analogous to the feature selection of relevant policies on the
Covid-19 epidemic studied in Ref.~\onlinecite{glielmo2022ranking}.
The authors of Ref.~\onlinecite{glielmo2022ranking} selected relevant policies for the number of deaths in the future among various conducted policies (e.g., school closing, stay-home restrictions, contact tracing, etc.).
They found that no {\it single} policy (feature) is sufficient on its own, yet the combination of {\it multiple} policies is getting effective in terms of predicting the number of future deaths.

\subsection{Feature selection by unsupervised learning}

Feature selection done in Sec.~\ref{sec:supervised} used the propensity data in a supervised learning manner. 
Here, we wish to perform feature selection without using propensity (or any dynamical data), which corresponds to an unsupervised learning method.

Again, the goal is to select features, constructing a feature vector, ${\bm X}_d$ in Eq.~(\ref{eq:selected_features}), 
yet without using the propensity.
Ideally, the selected features ${\bm X}_d$ and the original set ${\bm X}_{\rm Full}$ should be nearly equivalent, carrying essentially the same information, while $d \ll M$.
By using the language of information imbalance, this idea corresponds to $\Delta({\bm X}_{\rm Full} \to {\bm X}_d) \simeq \Delta({\bm X}_d \to {\bm X}_{\rm Full}) \simeq 0$.
To achive this condition we wish to minimize both $\Delta({\bm X}_{\rm Full} \to {\bm X}_d)$ and $\Delta({\bm X}_d \to {\bm X}_{\rm Full})$ simultaneously.
Thus, we define the symmetrized information
imbalance~\cite{glielmo2022ranking}, $\overline \Delta$, between ${\bm X}_{\rm Full}$ and ${\bm X}_d$ as  
\begin{equation}
    \overline \Delta ({\bm X}_{\rm Full}, \ {\bm X}_d)=\frac{\Delta({\bm X}_{\rm Full} \to {\bm X}_d) + \Delta({\bm X}_d \to {\bm X}_{\rm Full})}{\sqrt{2}} .
\end{equation}
We then minimize $\overline \Delta ({\bm X}_{\rm Full}, \ {\bm X}_d)$ using a greedy algorithm similar to the supervised learning case in Sec.~\ref{sec:supervised}.

First, among $M(=60)$ features, we search $X^{\rm 1st}$ at which the symmetrized information imbalance between ${\bm X}_{\rm Full}$ and ${\bm X}_1$ is minimized at $\overline \Delta ({\bm X}_{\rm Full}, \ {\bm X}_1)$, where ${\bm X}_1=(X^{\rm 1st})$.
We next search $X^{\rm 2nd}$ among the rest of $M-1$ features such that the symmetrized information imbalance is minimized at $\overline \Delta ({\bm X}_{\rm Full}, \ {\bm X}_2)$, where ${\bm X}_2=(X^{\rm 1st}, \ X^{\rm 2nd})$.
We repeat the above procedure until we find $X^{d{\rm th} }$ among the rest of $M-(d-1)$ features such that the symmetrized information imbalance is minimized at $\overline \Delta ({\bm X}_{\rm Full}, \ {\bm X}_d)$.

\begin{figure}
\includegraphics[width=0.98\columnwidth]{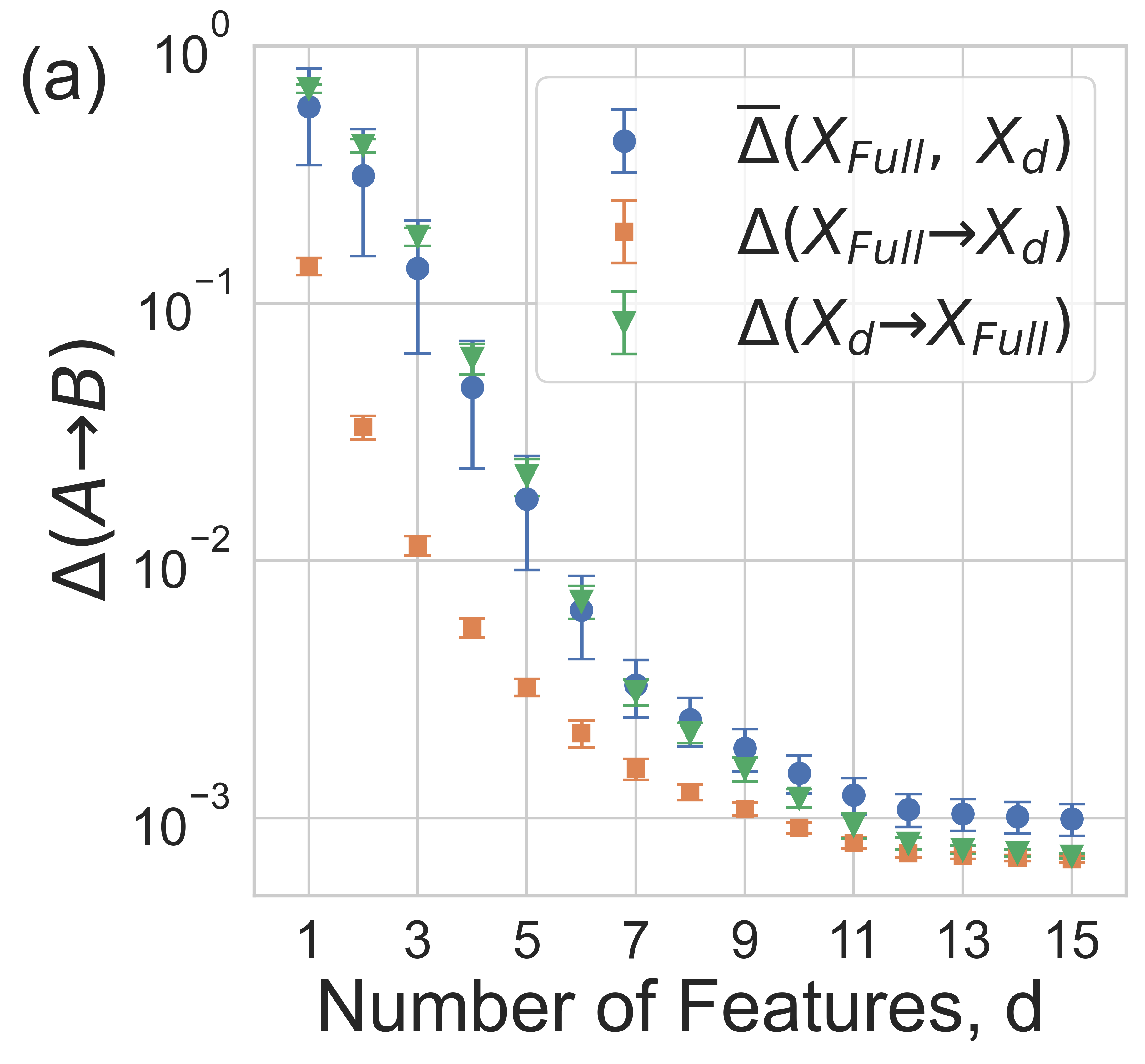}
\includegraphics[width=0.98\columnwidth]{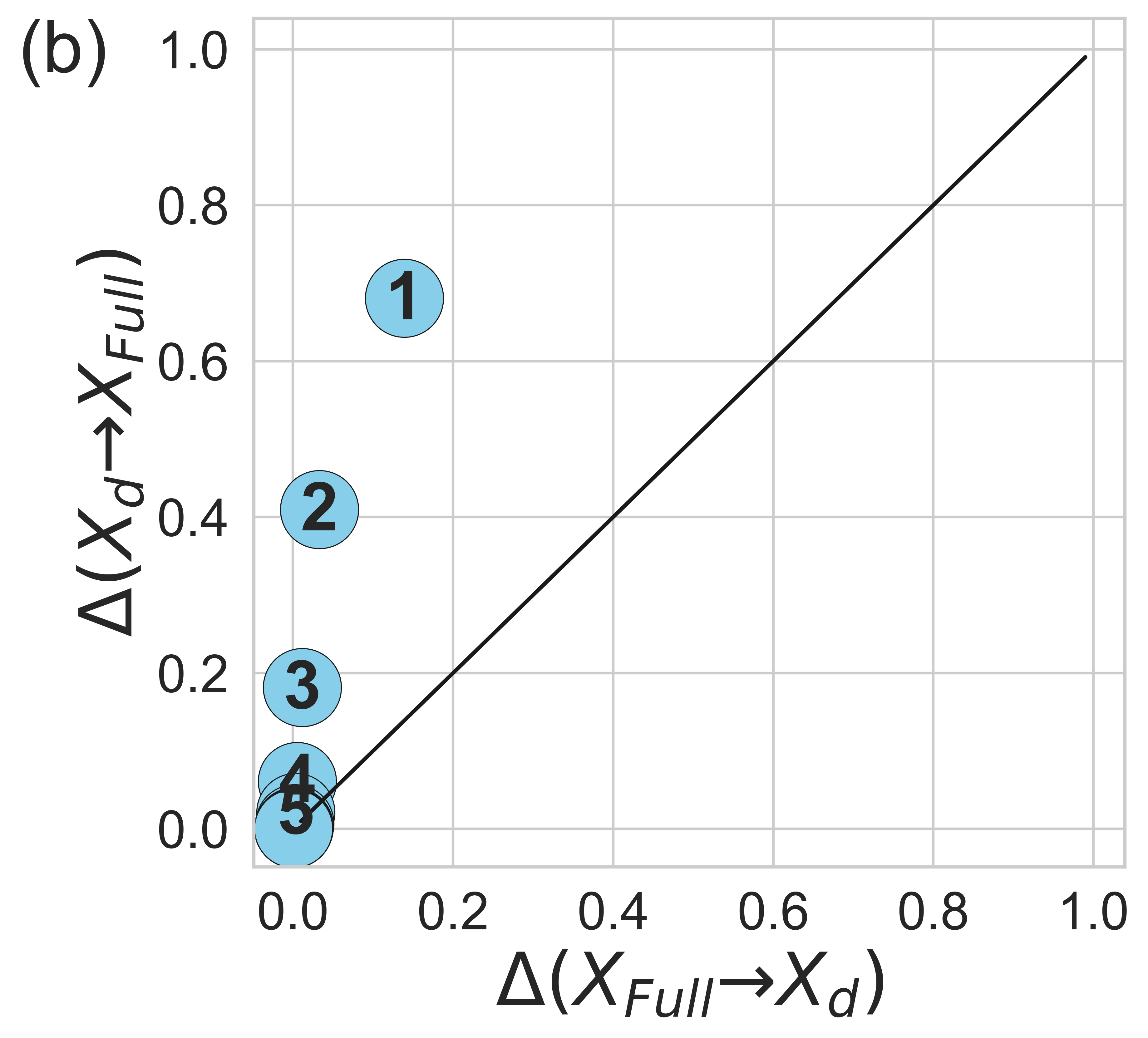}
\caption{Feature selection by the unsupervised approach. (a): Decay of the symmetric information imbalance, $\overline \Delta ({\bm X}_{\rm Full}, \ {\bm X}_d)$, as a function of $d$, together with $\Delta({\bm X}_{\rm Full} \to {\bm X}_d)$ and $\Delta({\bm X}_d \to {\bm X}_{\rm Full})$.
The errorbars correspond to the standard deviation from 10 different configurations.
(b): The same data for $\Delta({\bm X}_{\rm Full} \to {\bm X}_d)$ and $\Delta({\bm X}_d \to {\bm X}_{\rm Full})$ are presented in the information imbalance plance. The number inside a point represents $d$.}
\label{fig:IB_unsupervised}
\end{figure}

In Fig.~\ref{fig:IB_unsupervised}(a), we show the decay of $\overline \Delta ({\bm X}_{\rm Full}, \ {\bm X}_d)$ as a function of $d$, together with $\Delta({\bm X}_{\rm Full} \to {\bm X}_d)$ and $\Delta({\bm X}_d \to {\bm X}_{\rm Full})$.
We confirm that both of the terms, $\Delta({\bm X}_{\rm Full} \to {\bm X}_d)$ and $\Delta({\bm X}_d \to {\bm X}_{\rm Full})$, are minimized well (notice that the Y-axis is a log scale), yet decaying with a different speed, and show a saturation.
The corresponding information imbalance plane is shown in Fig.~\ref{fig:IB_unsupervised}(b). First, we find a strong asymmetry at $d=1$, $\Delta({\bm X}_{\rm Full} \to {\bm X}_1) \ll  \Delta({\bm X}_1 \to {\bm X}_{\rm Full})$, meaning that ${\bm X}_{\rm Full}$ is much more informative than ${\bm X}_1$, as expected.
Yet, with increasing $d$, symmetry is getting restored, converging to the origin, $(0, 0)$. 
The selected features up to $d=10$ from the 1st to $d$th, for the configuration presented in Fig.~\ref{fig:maps} are as follows: $\overline{\Theta_i}(\ell=4.5)$, $\overline{z_i}(\ell=3.5)$, $\overline{\Psi}_{6, i}(\ell =5)$, $\overline{\rho_i}(\ell=4.5)$,$\overline{\varphi_i}(\ell=0.5)$, $\overline{u_i}(\ell=4.5)$, $\overline{\varphi_i}(\ell=3.5)$, $\overline{u_i}(\ell =1.5)$, $\overline{\Theta_i}(\ell = 1)$, and  
$\overline{\rho_i}(\ell=1.5)$.
This lineup is quite similar to the one obtained by the supervised learning method in Sec.~\ref{sec:supervised}. Yet the order to achieve each lineup is very different. This difference will be highlighted in the machine learning prediction (see below).

\subsection{Application to machine learning glassy dynamics}

Having identified a reduced set of relevant structural features, ${\bm X}_d$,  from the original full set, ${\bm X}_{\rm Full}$, using information imbalance in both supervised and unsupervised manners, we now apply them to machine learning glassy dynamics to predict dynamic propensity.

The study of machine learning glassy dynamics was initiated by a series of papers by the group of Andrea Liu using a support vector machine that classifies mobile and immobile particles by making a hyperplane in a high-dimensional space of structural descriptors~\cite{cubuk2015identifying,schoenholz2016structural}.
This line of research was followed by neural network techniques, including cutting-edge graph neural networks, in order to increase the accuracy of prediction~\cite{bapst2020unveiling,jung2023predicting,shiba2023botan,pezzicoli2022se}. Typically, neural network methods outperform other methods, yet their necessity is still highly debated. In particular, linear regression with elaborated structural descriptors provides high accuracy as good as a deep learning method~\cite{alkemade2022comparing}.

Thus, in this study, we employ simple linear regression techniques.
To predict the dynamic propensity from the structural information, we construct a linear model,
\begin{equation}
    \hat p_i = {\bm w}^T {\bm X}_{d, i} + w^{(0)} ,
\label{eq:linear_model}
\end{equation}
where ${\bm w}=(w^{(1)}, \ w^{(2)}, \ \dots, \ w^{(d)})$ is the weight vector and $w^{(0)}$ is the intercept.
${\bm w}$ and $w^{(0)}$ are obtained by minimizing the mean squared error loss function, $\mathcal{L}^{\rm MSE} = \frac{1}{2N} \sum_{i=1}^N ( \hat p_i - p_i )^2$. 
%The hyperparameter tuning for the Lagrange multiplier is performed by the leave-one-out cross-validation.

In Fig.~\ref{fig:maps}(d), we show a map for predicted propensity $\hat p_i$, using the supervised scheme with $d=15$, for an unseen test configuration, which is the same configuration as Fig.~\ref{fig:maps}(a) where the ground truth $p_i$ are measured.  
Comparing with Fig.~\ref{fig:maps}(a), we visually confirmed that the linear model in Eq.~(\ref{eq:linear_model}) can predict the shape of dynamical heterogeneity qualitatively well. 
We also confirmed that the unsupervised method with $d=15$ performs prediction similarly (see below).

Besides, as a benchmark, we employ a supervised feature selection using a $L1$ norm penalty inspired by Lasso regression~\cite{bishop2006pattern}. In this method, we first minimize the loss function of Lasso regression, i.e., $\mathcal{L}^{\rm Lasso} = \mathcal{L}^{\rm MSE} + \frac{\alpha}{2} \sum_{f=1}^M|w^{(f)}|$, with a fixed strength of penalty $\alpha$. 
Because of the penalty term, some weights $w^{(f)}$ become strictly zero~\cite{bishop2006pattern}. Thus, the remaining features associated with non-zero weights can be considered relevant selected features. The number of selected features decreases with increasing $\alpha$. 
We tune the value of $\alpha$ to select $d$ features. 
We call this method of feature selection ``Lasso selection".
We then perform the standard linear regression (minimizing $\mathcal{L}^{\rm MSE}$ alone) by using $d$ selected features from the Lasso selection.

We also perform a random feature selection serving as a benchmark for the unsupervised method.
We randomly select $d$ features from the pool of $M$ features. For each configuration, we performed 1000 realizations and the subsequent linear regression.
We then averaged the results over them.
 
\begin{figure}
\includegraphics[width=0.98\columnwidth]{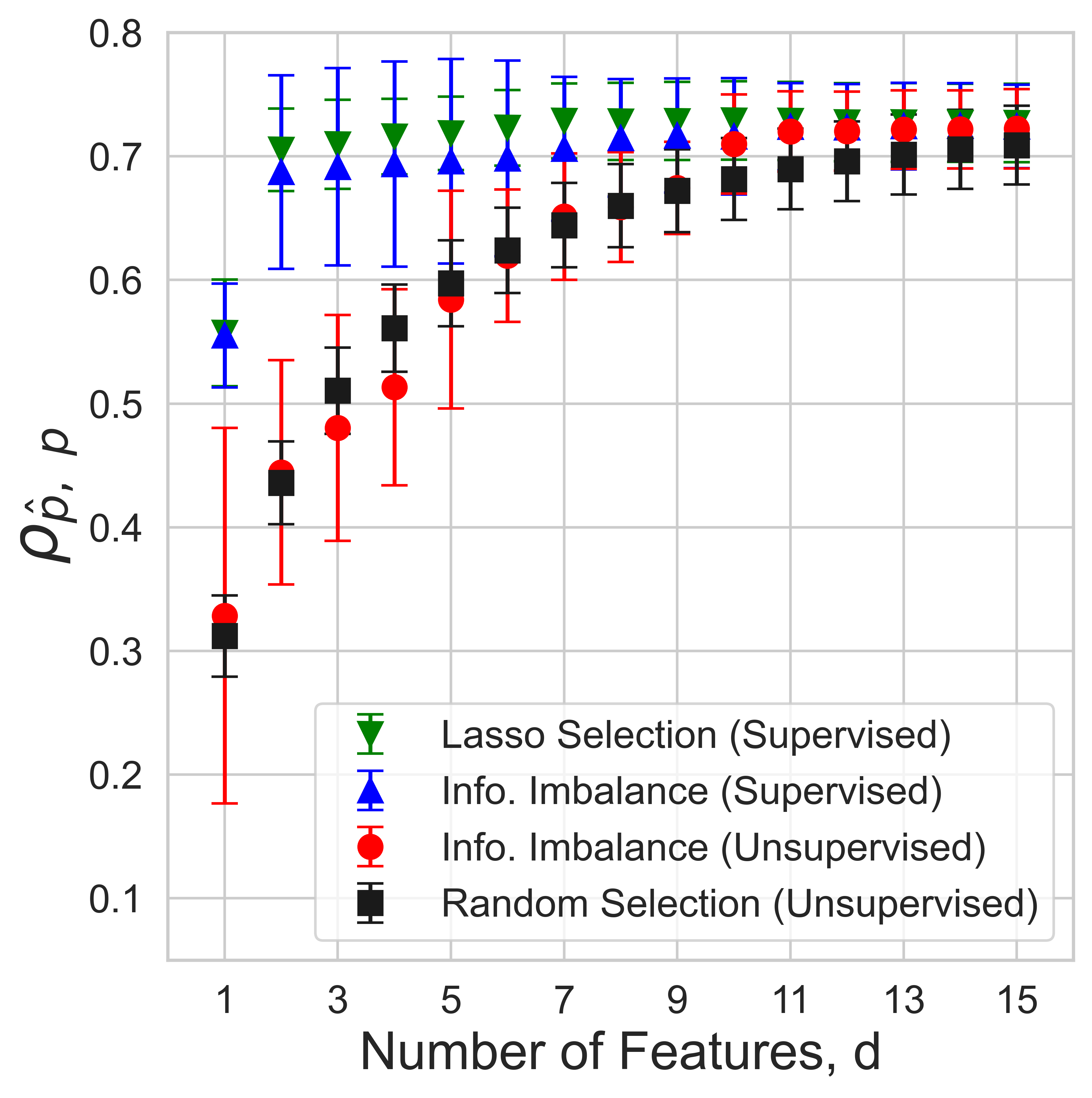}
\caption{ 
Pearson coefficient between the ground truth propensity $p_i$ at the structural relaxation time $\tau_\alpha \simeq 4 \times 10^6$ and the predicted one $\hat p_i$ by the linear regression using the selected $d$ features. Results from four different methods (see the main text in detail) are compared. The errorbars correspond to the standard deviation from 10 different (unseen) configurations.  
}
\label{fig:machine_learning_dynamics}
\end{figure}

Scatter plots comparing the ground truth $p_i$ and the prediction $\hat p_i$ are shown in Appendix~\ref{sec:scatter_plot}.
We then quantify the accuracy of the prediction by using the Pearson coefficient, $\rho_{p, \hat p}$, between $p_i$ and $\hat p_i$ for all feature selection methods we performed.
In Fig.~\ref{fig:machine_learning_dynamics}, we show $\rho_{p, \hat p}$ as a function of the number of selected features, $d$.
$\rho_{p, \hat{p}}$ is computed for all particles regardless of species. Species-wise values for $\rho_{p, \hat{p}}$ are reported in Appendix~\ref{sec:species_wise}.

Both of the supervised methods, the Lasso selection and information imbalance ones, quickly reach a high value, $\rho_{p, \hat p} \simeq 0.7$ after $d=2$, and seem to saturate, say, above $d=10$.
The first two selected features obtained by the supervised information imbalance method is typically $\overline \Theta_i(\ell)$ with $\ell \simeq 2-3$ and $\overline \rho_i(\ell)$ with $\ell \simeq 0.5-1.5$ , which is consistent with high Pearson coefficient with propensity observed in Fig.~\ref{fig:pearson_vs_IB}(a).
We wish to emphasize that the relative merit of the information imbalance method with respect to Lasso selection is that the former does not assume any model {\it a priori}, i.e., a non-parametric method, while
the latter assumes a specific model (the linear model in Eq.~(\ref{eq:linear_model})).

Turning our attention to the unsupervised methods, both the information imbalance and random feature selection methods converge slower than the supervised methods.
In particular, low values of $d$ result in a significant error in predicting the dynamics compared with supervised methods, and the unsupervised information imbalance method converges to similar performance as the supervised methods only beyond, say, $d = 10$. This is consistent with the fact that $d=10$ selected features by the supervised and unsupervised methods are quite similar. 
In contrast, random feature selection reaches a similar value only at the last data point, $d=15$, or slightly beyond, suggesting that the information imbalance more effectively selects useful features in the range of $d \simeq 10-15$ by comparing among features.

As we have seen, while the accuracy of the supervised information imbalance method jumps at $d=2$ directly to an approximate level of the large $d$ limit, the unsupervised one only gradually climbs up to that limit. Thus, the order of selection of each feature is quite different in the two methods, as we have confirmed.
This reflects that while the unsupervised information imbalance selects the $d$  features based on representing the best cloud of points of the full feature space, the dynamic propensity is best encoded in a different set of $d$ features that the supervised method manages to capture. 
Thus, the information imbalance techniques provide us with insight into the organization of structural and dynamical data points in high-dimensional feature space.

We focused on predicting glassy dynamics on the timescale of structural $\alpha$ relaxation, using the propensity, $p_i(t)$, measured at $t=\tau_{\alpha} \simeq 4 \times 10^6$. Additionally, we made predictions for a shorter timescale, $t=5 \times 10^4$, which is close to $\beta$ relaxation timescales, as described in Appendix~\ref{sec:shorter_timescale}. Although the Pearson coefficient values are generally lower, consistent with previous studies~\cite{jung2023roadmap}, the overall trend remains similar to that observed for the $\alpha$ relaxation timescale.

\section{Discussion and Conclusion}
\label{sec:discusssion_conclusion}

We applied the information imbalance techniques proposed recently~\cite{glielmo2022ranking} to select relevant structural features in glassy dynamics among a pool of various physically motivated descriptors with different coarse-graining lengthscales.
Unlike the conventional statistical measures characterizing the relationship between variables $A$ and $B$, such as the Pearson coefficient and mutual information, the information imbalance is constructed in an asymmetric manner, which allows us to ask not only if $A$ and $B$ are equivalent or independent, but also if
one is more informative than the other, without referring to other variables.

We confirmed that the information imbalance between dynamic propensity and structural features can capture the general trend obtained by the Pearson coefficient, including its dependence on the coarse-graining lengthscale. We then performed feature selection from the full set of structural features in a supervised manner using the propensity data.
We found that although some descriptors (e.g., the steric order parameter) have higher correlations with propensity, no {\it single} descriptor is more informative than propensity, at least within our pool of descriptors. Instead, a set of selected {\it multiple} features is more informative than propensity, showing a strong asymmetry in the information imbalance plane.
We also performed feature selection in an unsupervised manner without using any dynamics data. 
Remarkably, although the order of selection of each feature is quite different at smaller $d<10$, the selected features at $d=10$ are similar to those obtained by the supervised method using propensity.
We confirmed that selected structural features in both supervised and unsupervised manners are indeed highly relevant and useful for predicting future glassy dynamics, comparing with other feature selection techniques.

This work corresponds to proof of the concept for the newly introduced idea of information imbalance. Thus, so far, we have focused only on one simulation model, one temperature (lowest temperature), and two timescales (structural $\alpha$ relaxation timescale in the main text and a $\beta$ relaxation timescale in Appendix~\ref{sec:shorter_timescale}). 
It is interesting to explore and compare different models (2D vs. 3D.~\cite{flenner2015fundamental}, fragile vs. strong~\cite{furukawa2016significant,kim2013multiple} etc.), different temperatures, and different timescales (including much lower temperatures and longer timescales~\cite{scalliet2022thirty}). It is interesting to see if relevant descriptors and optimal coarse-graining lengthscale change depending on systems, the state point, and the timescale.

Besides, we studied only simple, basic physical descriptors, such as local energy and the steric order parameter. 
A wide variety of physical descriptors characterizing structure have been proposed, such as locally favored structures~\cite{coslovich2007understanding,royall2015role}, vibrational modes~\cite{widmer2008irreversible,manning2011vibrational}, softness~\cite{schoenholz2016structural,ridout2024dynamics}, caging potential~\cite{sahu2024structural,anwar2024exploring}, quasilocalized excitations~\cite{richard2023detecting}, mesoscopic elastic stiffness~\cite{kapteijns2021does}, local yield stress~\cite{barbot2018local,lerbinger2022relevance}, overlap field~\cite{charbonneau2016linking,guiselin2022statistical}, local configurational entropy~\cite{berthier2021self}, topological defects~\cite{baggioli2021plasticity,wu2023topology}, and so on (see also Ref.~\onlinecite{richard2020predicting} for a review).
It would be insightful to include these descriptors in the information imbalance analysis to identify and select the most relevant features from the wide variety of proposed descriptors.

%\AS{Can we also include the following result: In the supervised Information Imbalance, we showed that the physical descriptors contain information about the dynamics. But is it coming to a saturation value and not decreasing after 13?? descriptors and not reaching zero  (Ideal case where the descriptors contain complete information about propensity). It implies that present descriptors can only partially predict the dynamical space, and there's an open space for undiscovered descriptors that can explain dynamics, or it's not possible to predict the dynamics from physical descriptions. }
%\MO{This is nice idea. This takes into account the sentence below.}

The above exercise would also be related to identifying the correct theory of glass transition. Many theories have been proposed based on quite distinct physical mechanisms, such as entropy~\cite{lubchenko2007theory,bouchaud2004adam}, elasticity~\cite{lemaitre2014structural,ozawa2023elasticity,tahaei2023scaling,dyre2024solid}, locally preferred structures~\cite{tarjus2005frustration,turci2017nonequilibrium}, etc.
Information imbalance may help to identify the relevant physical mechanism associated with the selected feature.
Furthermore, it would be worthwhile to investigate whether some of the descriptors mentioned above (and undiscovered ones) are more informative than dynamic propensity. Another scenario would be that none of the above features has a dominant role in dynamics. Instead, a combination of them becomes informative in terms of glassy dynamics. If this is the case, one cannot attribute a {\it single} dominant physical mechanism for glassy dynamics. Instead, {\it multiple} mechanisms would be hybridized in reality.
Yet, there is also a possibility that structural data alone cannot contain complete information about dynamics~\cite{chandler2010dynamics}.
The information imbalance techniques can be useful in assessing these different scenarios quantitatively.

\begin{acknowledgments}

We thank Ali Hassanali for introducing us to the information imbalance analysis. We also thank Gerhard Jung, Takumi Nagasawa, Rob Jack, and Daniele Coslovich for their discussions and suggestions.
This work was supported by the Agence Nationale de la Recherche (ANR-19-P3IA-0003) under the Plan France 2030.
\end{acknowledgments}

\section*{Data Availability}

All the source codes and dataset used in this paper are openly available at
\href{https://github.com/Anand-Sharma22/Feature-selection-Information-Imbalance}
{https://github.com/Anand-Sharma22/Feature-selection-Information-Imbalance}.

\appendix

\section{Scatter plots for propensity}
\label{sec:scatter_plot}

\begin{figure}
\includegraphics[width=0.48\columnwidth]{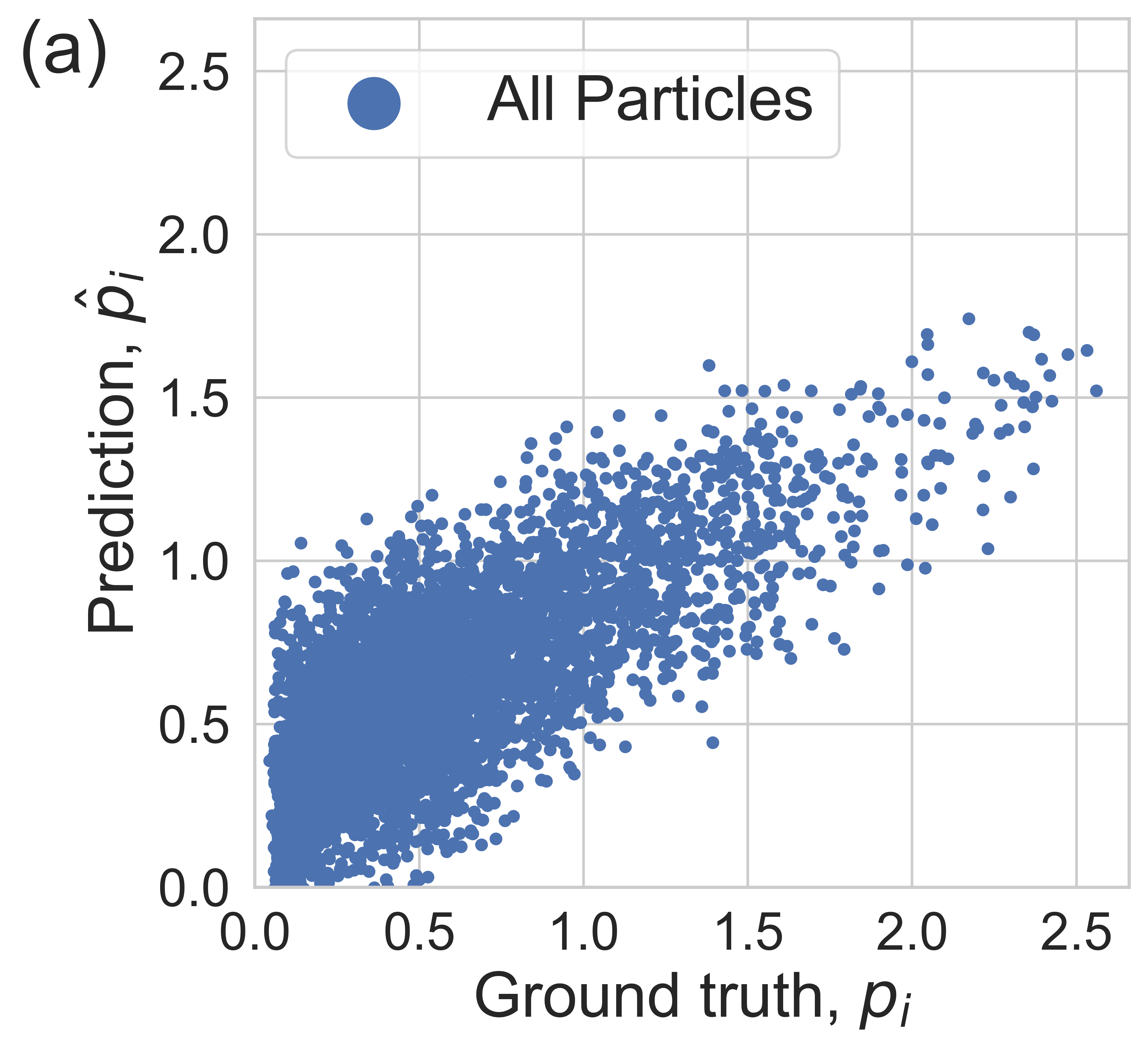}
\includegraphics[width=0.48\columnwidth]{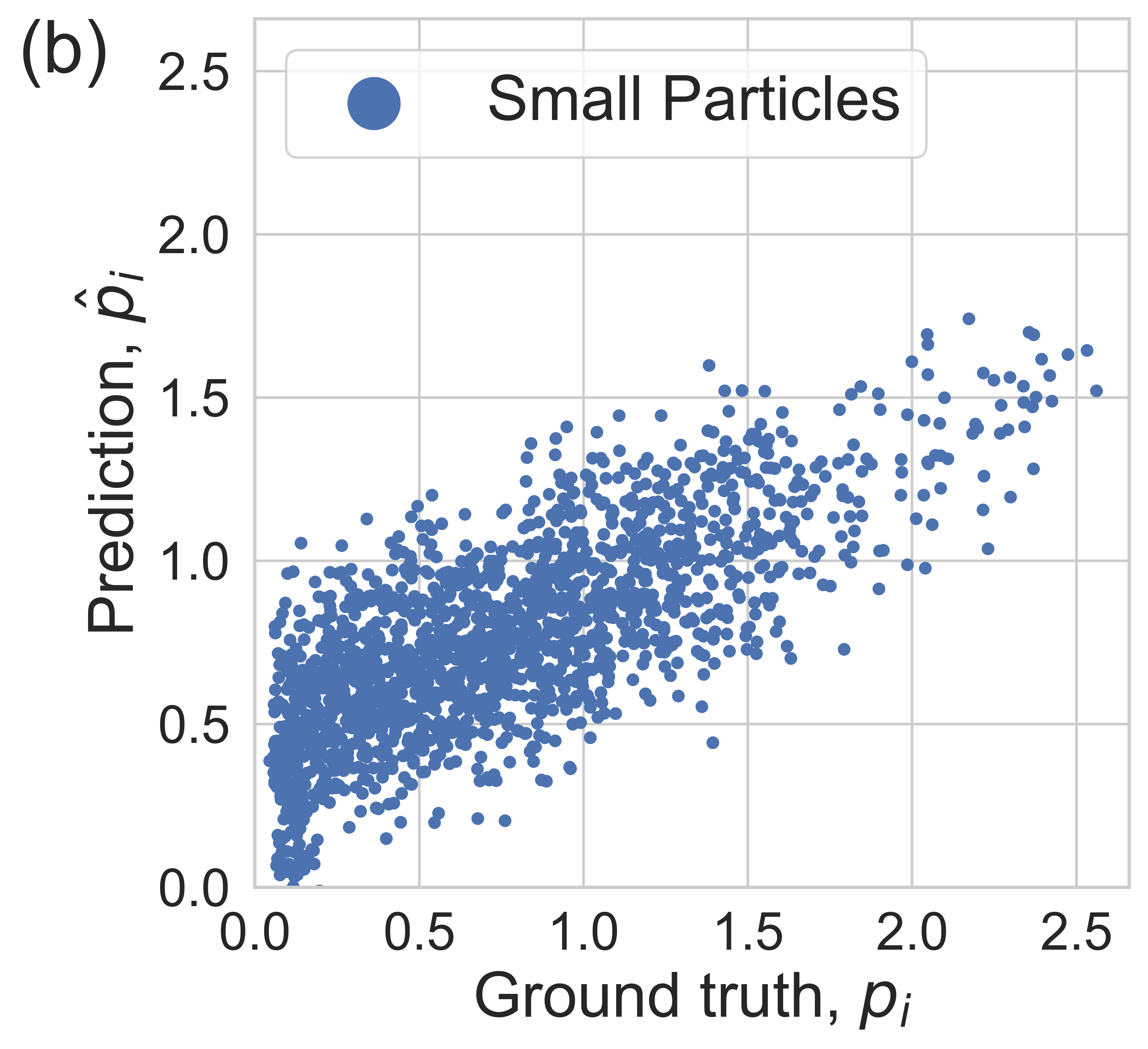}
\includegraphics[width=0.48\columnwidth]{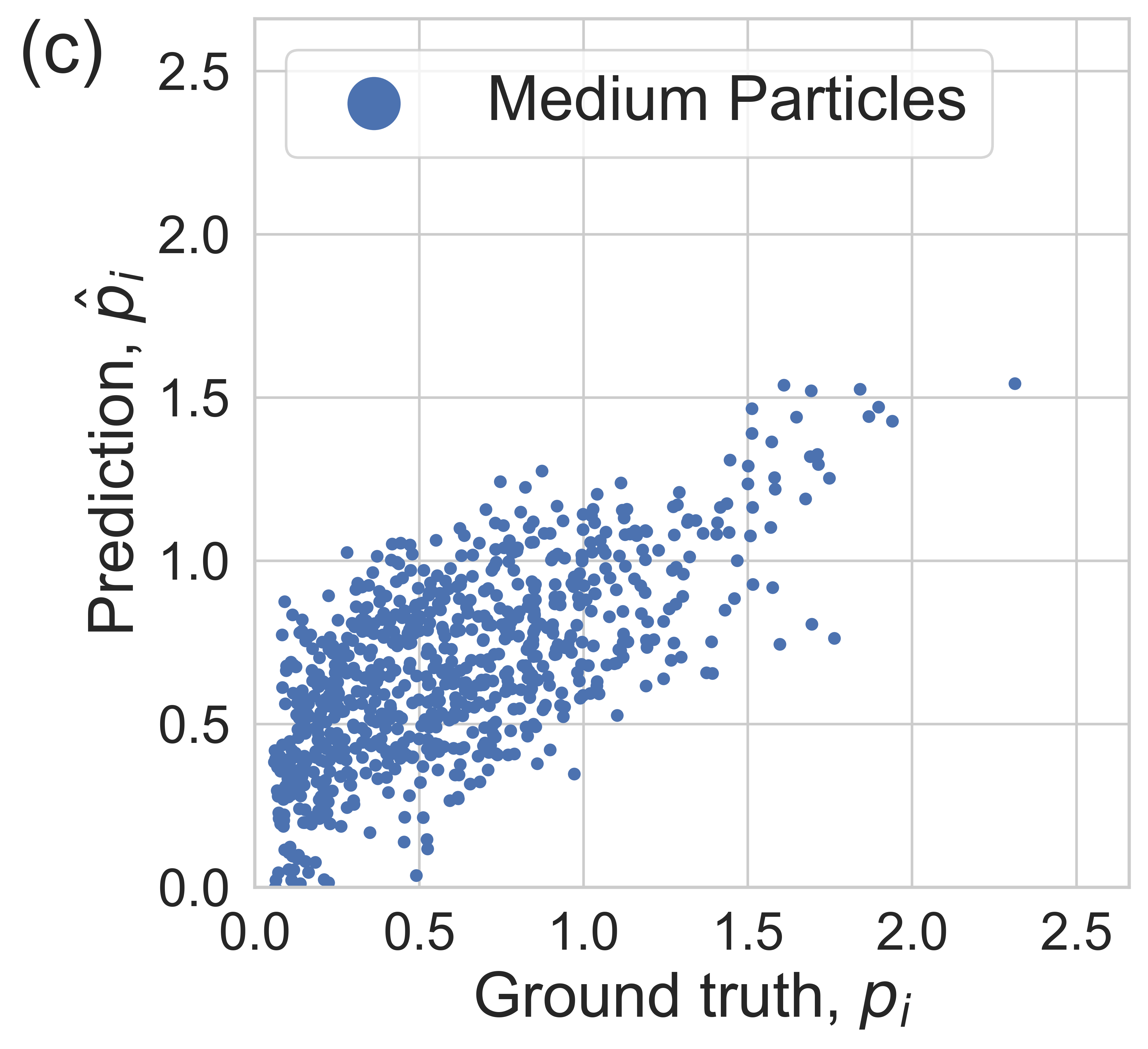}
\includegraphics[width=0.48\columnwidth]{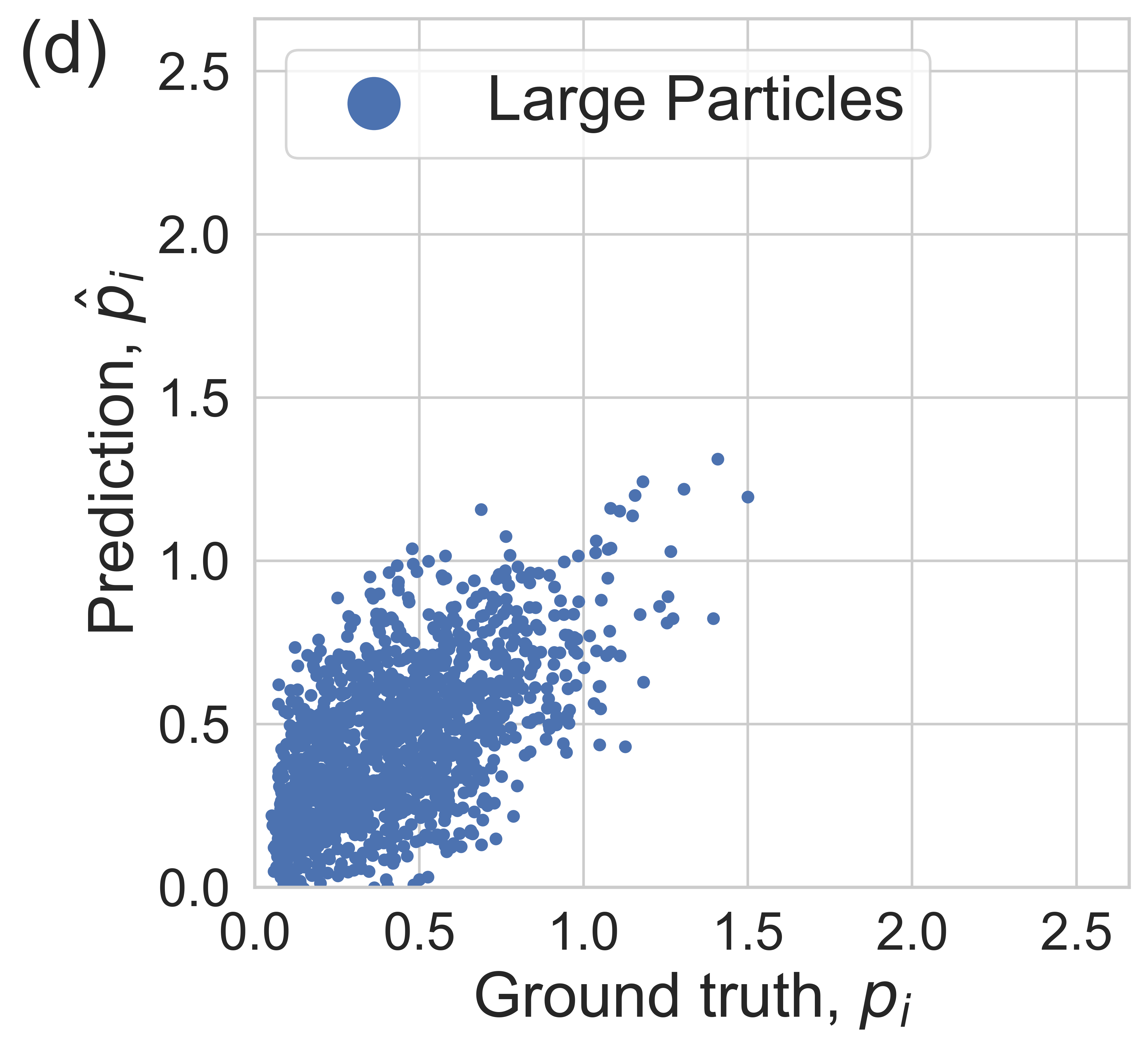}
\caption{ 
Scatter plots comparing the ground truth propensity $p_i$ with the predicted values $\hat p_i$ for all particles (a), small particles (b), medium particles (c), and large particles (d), obtained using linear regression with $d=15$ features selected by the supervised information imbalance method.
}
\label{fig:scatter_plots}
\end{figure}

In Fig.~\ref{fig:scatter_plots}, we present scatter plots comparing the ground truth propensity $p_i$ with the predicted values $\hat p_i$, obtained using linear regression with $d=15$ features selected by the supervised information imbalance method described in Sec.~\ref{sec:supervised}. These plots correspond to the configuration shown in Fig.~\ref{fig:maps} as an example. Figure~\ref{fig:scatter_plots}(a) displays the scatter plot for all particles, while Figs.~\ref{fig:scatter_plots}(b, c, d) show the scatter plots for small (b), medium (c), and large (d) particles, respectively.
We note that in this paper, training (minimizing the loss function) is performed simultaneously for all particles for simplicity. Training each species separately could improve prediction performance~\cite{jung2023roadmap}.

The plot for smaller particles shows a wider range of propensity, as expected, while the data for medium and large particles span a narrower range. Overall, these plots exhibit a standard oval shape without skewness, indicating that the Pearson coefficient can appropriately quantify the degree of correlation.

\section{Species-wise plots}
\label{sec:species_wise}

\begin{figure*}
\includegraphics[width=0.66\columnwidth]{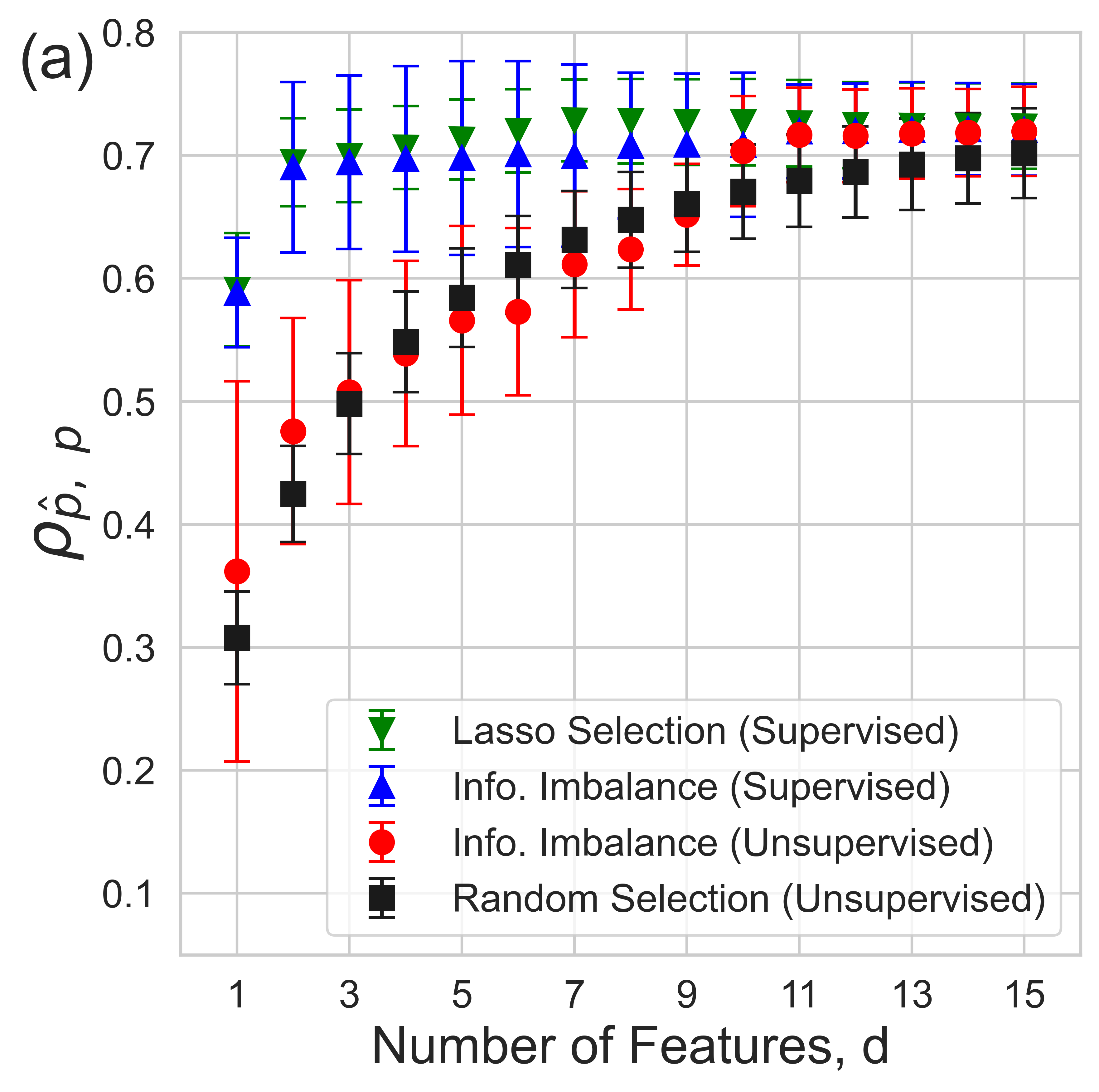}
\includegraphics[width=0.66\columnwidth]{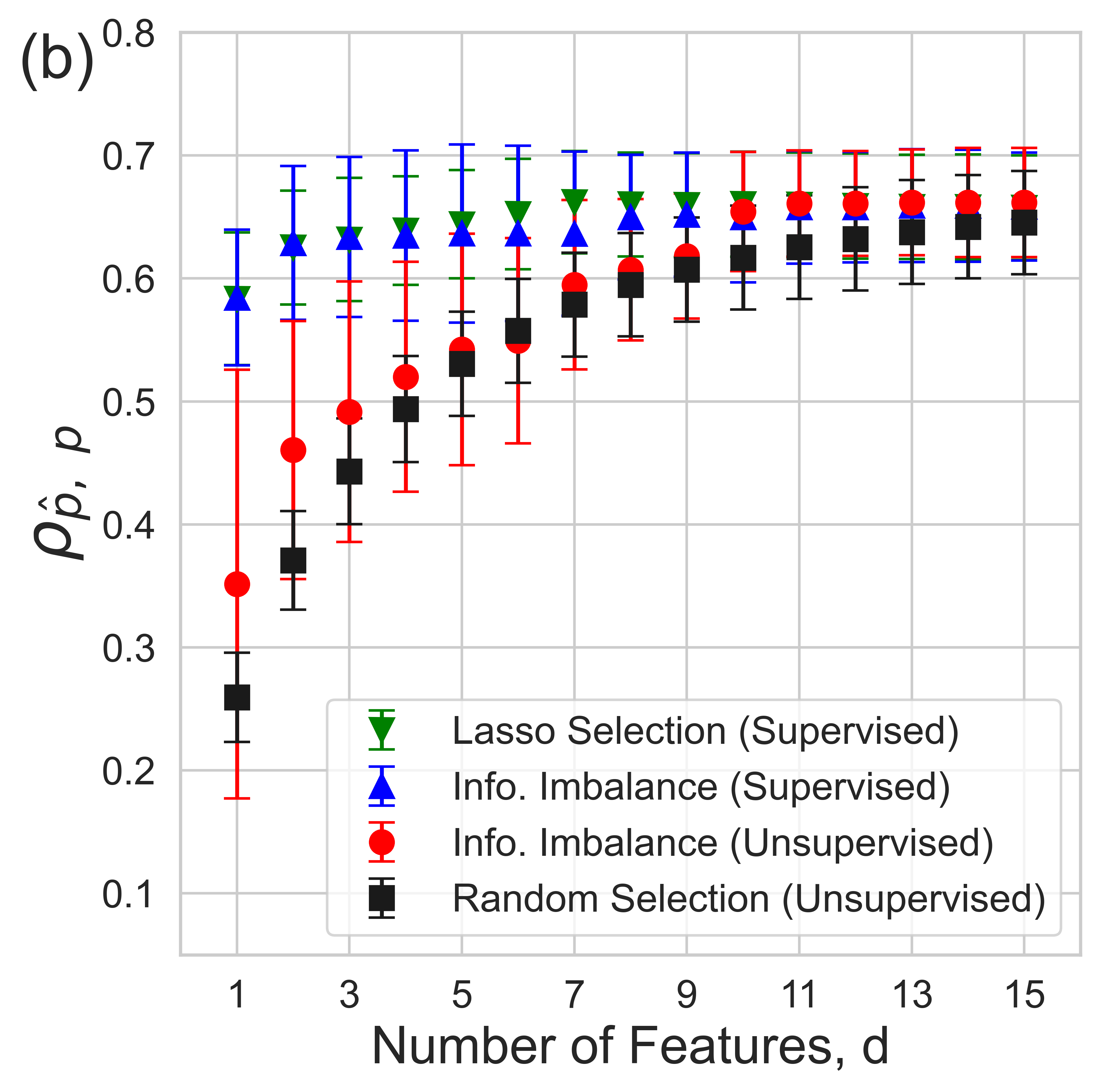}
\includegraphics[width=0.66\columnwidth]{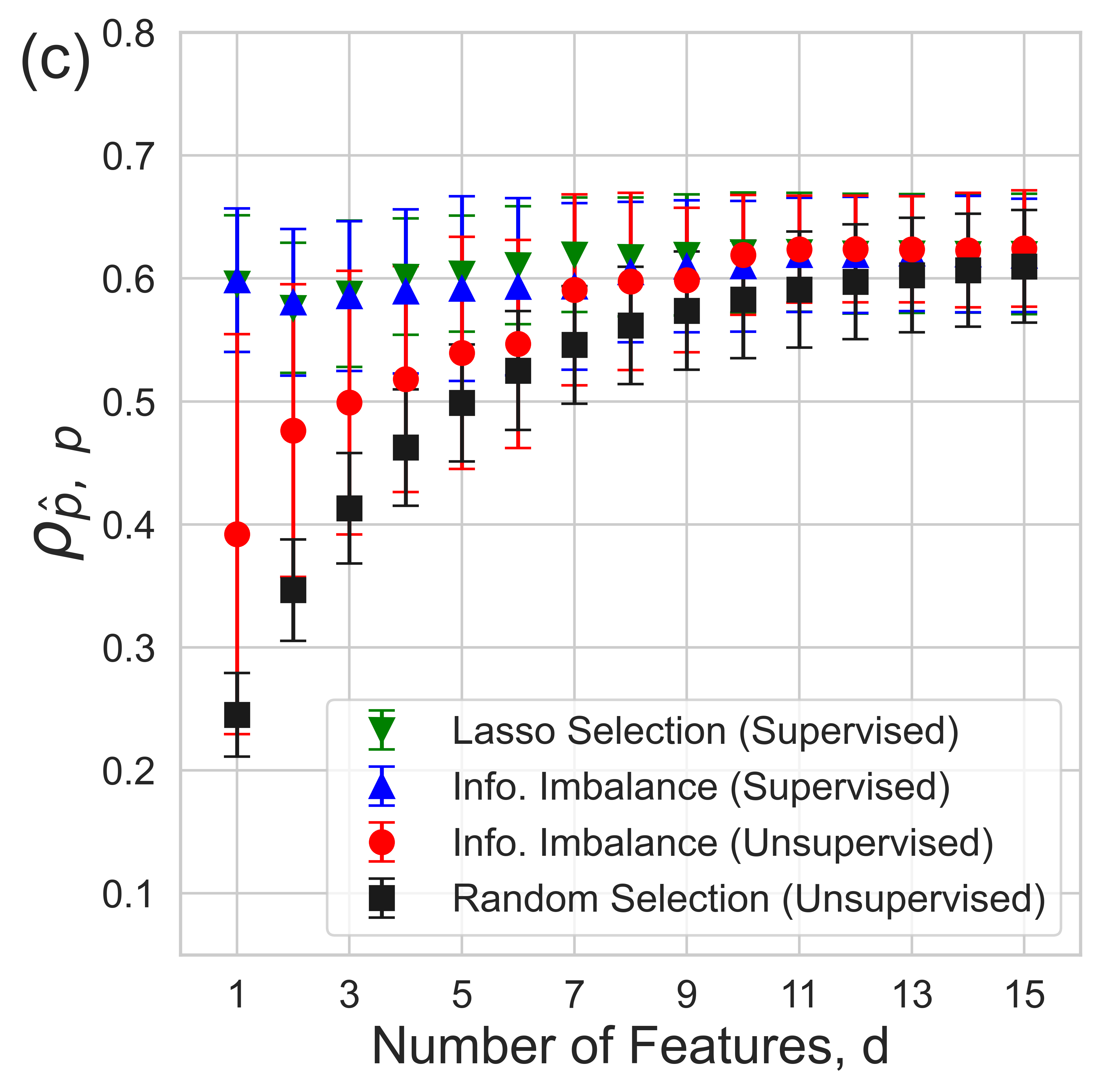}
\caption{ 
Pearson coefficient between the ground truth propensity $p_i$ at the structural relaxation time $\tau_\alpha \simeq 4 \times 10^6$ and the predicted propensity $\hat p_i$, obtained from linear regression using the selected $d$ features. The figure format follows that of Fig.~\ref{fig:machine_learning_dynamics}, but the Pearson coefficients are computed separately for each species: small (a), medium (b), and large (c) particles, respectively.
}
\label{fig:species_wise}
\end{figure*}

As anticipated from Fig.~\ref{fig:scatter_plots}, the Pearson coefficient values differ across species. In Fig.~\ref{fig:species_wise}, we present the species-wise Pearson coefficient, calculated separately for small, medium, and large particles. The value is highest for small particles, while medium and large particles show lower values, with the coefficient decreasing as particle size increases. Nevertheless, the overall trend remains consistent with the Pearson coefficient computed for all particles combined across species in Fig.~\ref{fig:machine_learning_dynamics}.

\section{Shorter timescale}
\label{sec:shorter_timescale}

In the main text, we studied the machine learning prediction of the propensity measured at the structural $\alpha$ relaxation time for $T=0.30$. We also study a short timescale, $t=5 \times 10^4$, close to $\beta$ relaxation timescales of the same temperature (see Fig.~\ref{fig:Fskt}).

\begin{figure}
\includegraphics[width=0.98\columnwidth]{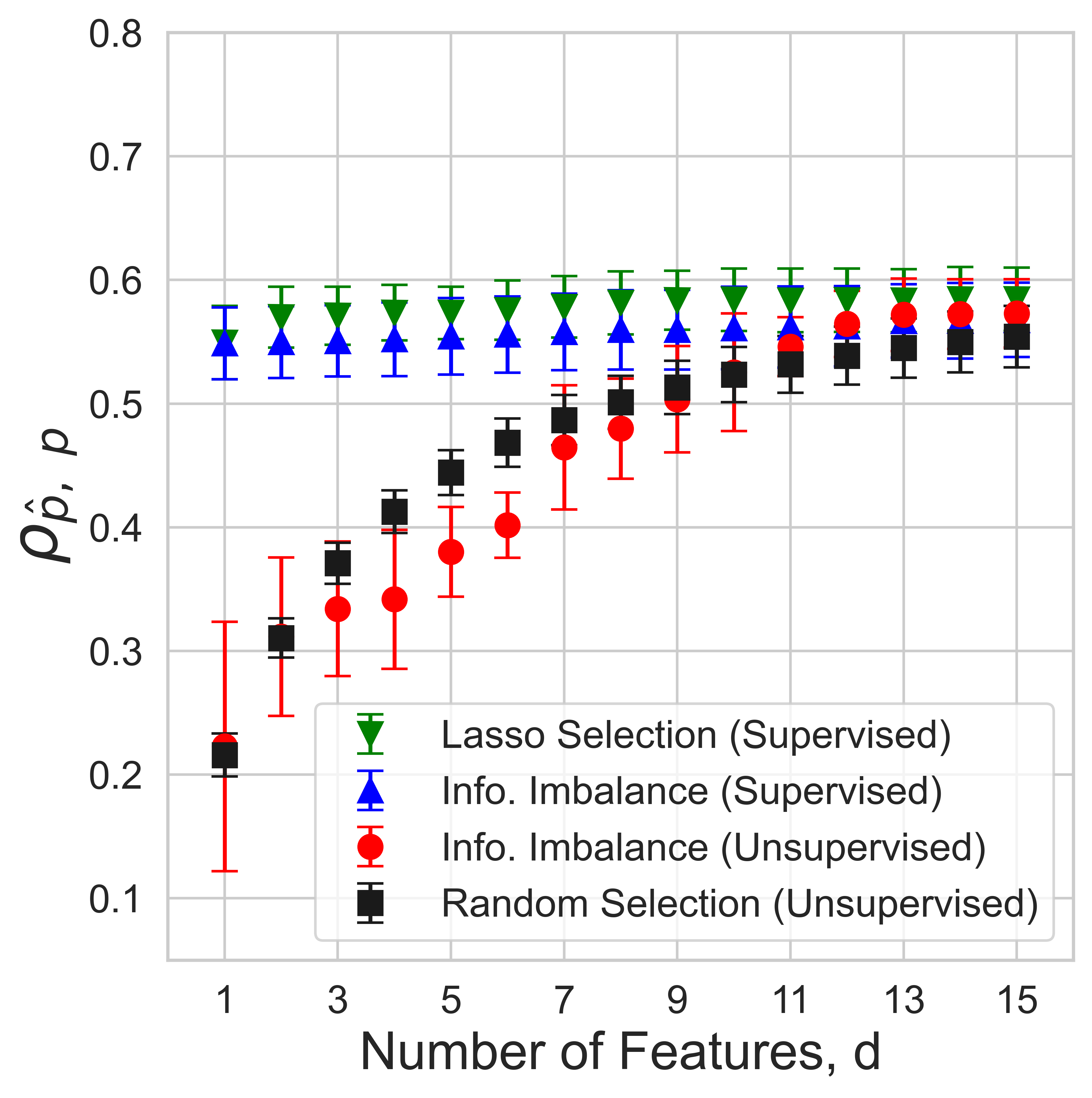}
\caption{ 
Pearson coefficient between the ground truth propensity $p_i$ at a shorter time $t = 5 \times 10^4$ (near $\beta$ relaxation timescales) and the predicted one $\hat p_i$ by the linear regression using the selected $d$ features. The figure format follows that of Fig.~\ref{fig:machine_learning_dynamics}.
}
\label{fig:machine_learning_dynamics_short}
\end{figure}

Figure~\ref{fig:machine_learning_dynamics_short} presents the Pearson coefficient obtained using the four feature selection methods we employed. We found that the Pearson coefficient at the shorter timescale is lower than at the structural relaxation time, consistent with previous reports~\cite{jung2023roadmap}. Additionally, for the supervised methods, the value at $d=1$ is already close to the asymptotic value at large $d$. The feature identified at $d=1$ by the supervised information imbalance method is $\overline \Theta_i(\ell)$ with $\ell \simeq 1$, corresponding to a slightly smaller coarse-graining length scale than at the $\alpha$ relaxation timescale, which aligns with previous findings~\cite{tong2018revealing}.

We also observed that the unsupervised information imbalance method underperforms compared to the random selection method around $d=5$, but outperforms it above $d=10$. This suggests that the features selected by the unsupervised information imbalance method (which rely solely on static information) near $d=5$ are not relevant for short-time dynamics.

%\nocite{*}
\bibliography{references}% Produces the bibliography via BibTeX.

\end{document}